\documentclass[a4paper]{jpconf}
\usepackage{graphicx}

\usepackage{amssymb}
\usepackage{epsf,epsfig,psfrag}

\def\be{\begin{equation}}
\def\ee{\end{equation}}
\def\bea{\begin{eqnarray}}
\def\eea{\end{eqnarray}}

\begin{document}
\title{Neutrino oscillations: theory and phenomenology \footnote{
Talk given at the XXII International Conference on Neutrino 
Physics and Astrophysics ``Neutrino 2006'', Santa Fe, June 13-19, 
2006}}

\author{E K Akhmedov \footnote{
On leave from the National Research Center ``Kurchatov Institute'', 
Moscow, Russia }
}

\address{Department of Theoretical Physics, Royal Institute of 
Technology, AlbaNova University Center, SE-106 91 Stockholm, Sweden }

\ead{akhmedov@ictp.trieste.it}

\begin{abstract}
A brief overview of selected topics in the theory and phenomenology of 
neutrino oscillations is given. These include: oscillations in vacuum and 
in matter; phenomenology of 3-flavour neutrino oscillations and effective 
2-flavour approximations; CP and T violation in neutrino oscillations in 
vacuum and in matter; matter effects on $\nu_\mu \leftrightarrow \nu_\tau$ 
oscillations; parametric resonance in neutrino oscillations inside the 
earth; oscillations below and above the MSW resonance; unsettled issues in 
the theory of neutrino oscillations.

\end{abstract}

\section{A bit of history...}

The idea of neutrino oscillations was first put forward by Pontecorvo in 
1957 \cite{Pont}. Pontecorvo suggested the possibility of $\nu\leftrightarrow 
\bar{\nu}$ oscillations, by analogy with ~$K^0\bar{K}^0$ oscillations
(only one neutrino species -- $\nu_e$ -- was known at that time). 
Soon after the discovery of muon neutrino, Maki, Nakagawa and Sakata 
\cite{MNS} suggested the possibility of neutrino flavour transitions (which 
they called ``virtual transmutations'').

\begin{figure}[h]   
\hbox{\hfill
\hspace*{1.8cm}
{\includegraphics[height=3.0cm,angle=90]{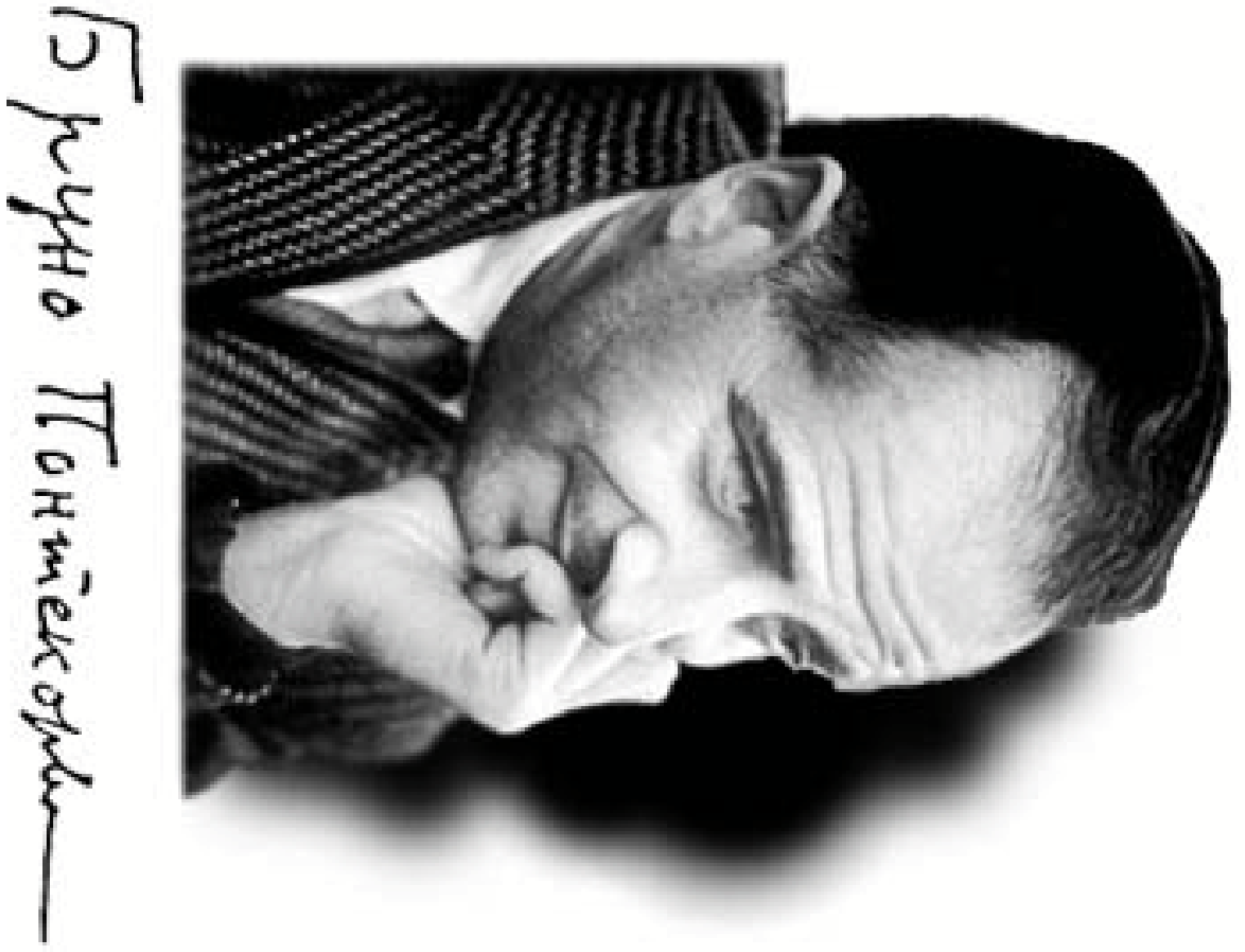}}
{\includegraphics[height=6.0cm,angle=90]{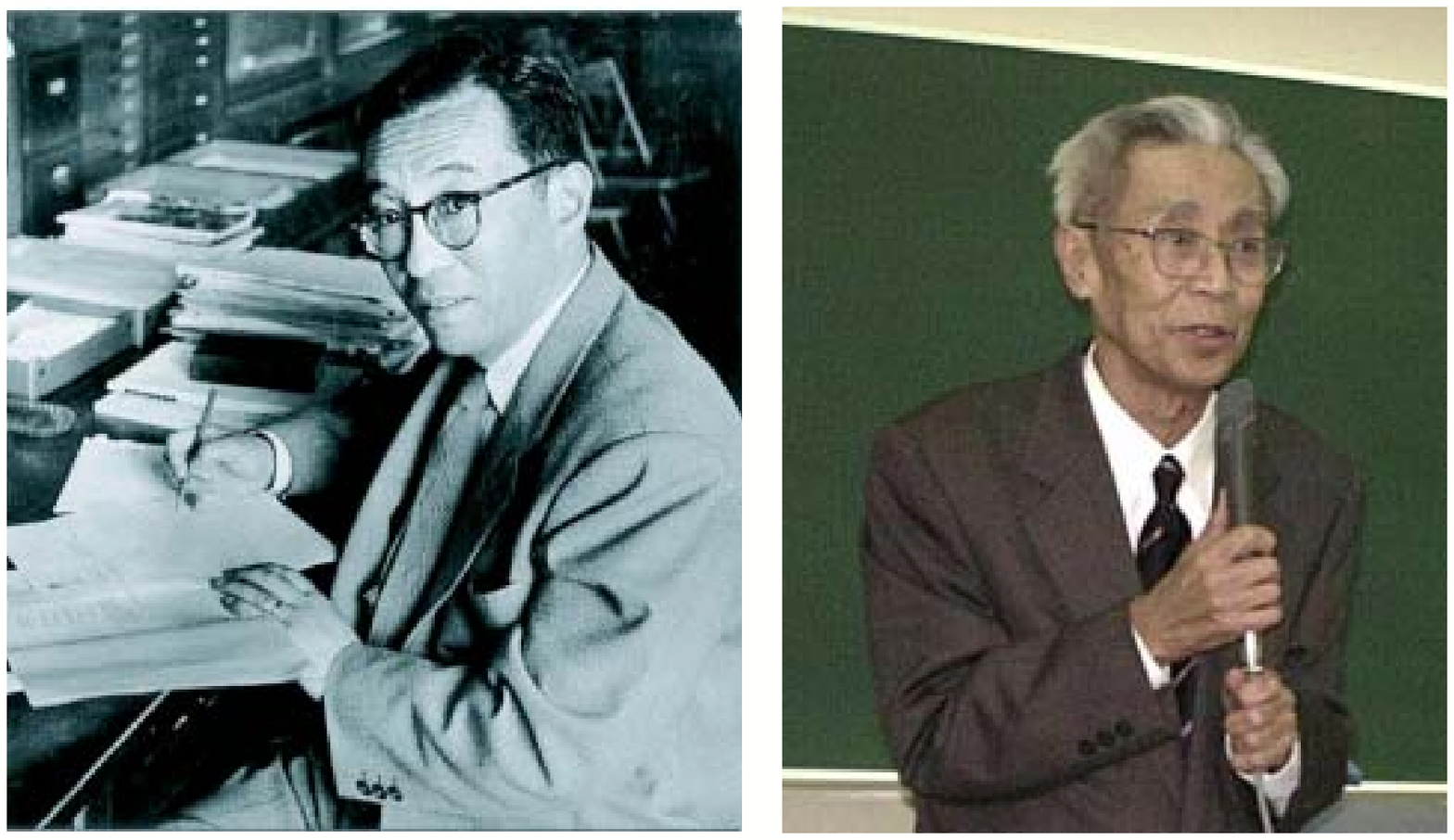}}
{\includegraphics[height=3.43cm]{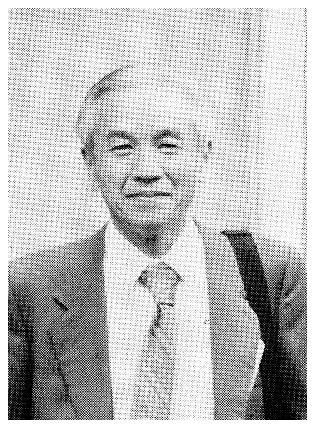}}
\hfill}
\label{founders}
\caption{Bruno Pontecorvo (1913 - 1993), Shoichi Sakata 
(1911 - 1970), Ziro Maki (1929 -- 2005) and  Masami  Nakagawa (1932 - 2001). 
}
\end{figure}

\section{Theory}
\subsection{Neutrino oscillations in vacuum}   
Neutrino oscillations are a manifestation of leptonic mixing. For massive 
neutrinos, weak (flavour) eigenstates do not in general coincide with mass 
eigenstates but are their linear combinations; the diagonalization of the 
leptonic mass matrices leads to the emergence of the leptonic mixing matrix 
in the expression for the charged current interactions, the relevant part of 
the leptonic Lagrangian being 
\be
 -{\cal L}_{w+m}~=~\frac{g}{\sqrt{2}}(\bar{e}_L\gamma_\mu 
\,V_L^\dag U_L\, \nu_L)\,W^\mu~+~{\rm diag.~~mass~~terms}\,.
\ee
Here $V_L$ and $U_L$ are the left-handed unitary transformations  
that diagonalize the mass matrices of charged leptons and neutrinos. 
The leptonic mixing matrix $U=V_L^\dag U_L$ relates the 
left-handed components of the neutrino mass eigenstates 
and flavour eigenstates according to 
\be
|\nu^{\rm fl}_a\rangle = \sum_i U_{ai}^*\,|\nu^{\rm mass}_i\rangle
\qquad(a=e,\mu,\tau;~i=1,2,3)\,.
\ee
For relativistic neutrinos, the oscillation probability in vacuum is   
\be
P(\nu_a\to\nu_b; L) = \left|\sum_i U_{bi}\; e^{-i \frac{m_i^2}{2p} L} \;
U_{ai}^*\right|^2\,. 
\ee
For 2-flavour (2f) oscillations, which are a good first approximation in 
many cases, one has 
\be
|\nu_e\rangle ~=~ \cos\theta\, |\nu_1\rangle +\sin\theta\, |\nu_2\rangle
\,,\,\,\,
\ee
\be
|\nu_\mu\rangle \,=\,-\sin\theta\,|\nu_1\rangle +\cos\theta\,|\nu_2\rangle\,,
\ee
and eq.~(1) yields the 2f transition probability 
\be
P_{\rm tr} = \sin^2 2\theta \sin^2 \left(\frac{\Delta m^2}{4E}\,L\right)\,.
\ee
The modes of neutrinos oscillations depend on the character of neutrino 
mass terms:
\begin{itemize}

\item Dirac mass terms ($\bar{\nu}_L m_D N_R + h.c.$):  
active - active oscillations 
$\nu_{aL}~\leftrightarrow~ \nu_{bL}$ 
 $(a,b = e,\,\mu,\,\tau)$ \\
~ Neutrinos are Dirac particles.

\item Majorana mass terms ($\bar{\nu}_L m_L (\nu_L)^c + h.c.$): active - 
active oscillations $\nu_{aL}~\leftrightarrow~ \nu_{bL}$. \\
Neutrinos are Majorana particles. 

\item Dirac~+~Majorana mass terms ($\bar{\nu}_L m_D N_R + \bar{\nu}_L m_L 
(\nu_L)^c + \bar{N}_R M (N_R)^c + h.c.$): active - active oscillations 
$\nu_{aL}~\leftrightarrow \nu_{bL}$;  active - sterile oscillations 
$\nu_{aL}~\leftrightarrow~ (N_{bR})^c \equiv (N^c_b)_{L}$. \\
Neutrinos are Majorana particles. 

\end{itemize}
Would an observation of active - sterile neutrino oscillations mean that 
neutrinos are Majorana particles? Not necessarily! In principle, one can 
have active - sterile oscillations with only Dirac - type mass terms at the 
expense of introducing additional species of sterile neutrinos with opposite 
lepton number $L$.

\subsection{Neutrino oscillations in matter -- The ~MSW ~effect \cite{MSW}} 

Matter can change the pattern of neutrino oscillations drastically. 
In particular, a resonance enhancement of oscillations and resonance flavour 
conversion become possible (Wolfenstein, 1978; Mikheyev \& Smirnov, 
1985 \cite{MSW}). 
\begin{figure}[h]   
\hbox{\hfill
\hspace*{2.3cm}
\includegraphics[height=3.0cm]{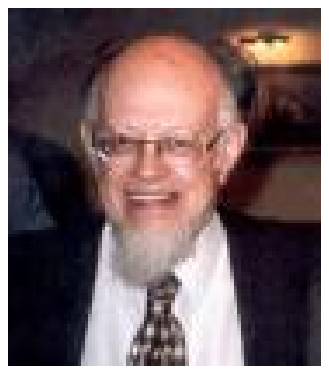}
\hspace*{0.3cm}
\includegraphics[height=5.25cm,angle=90]{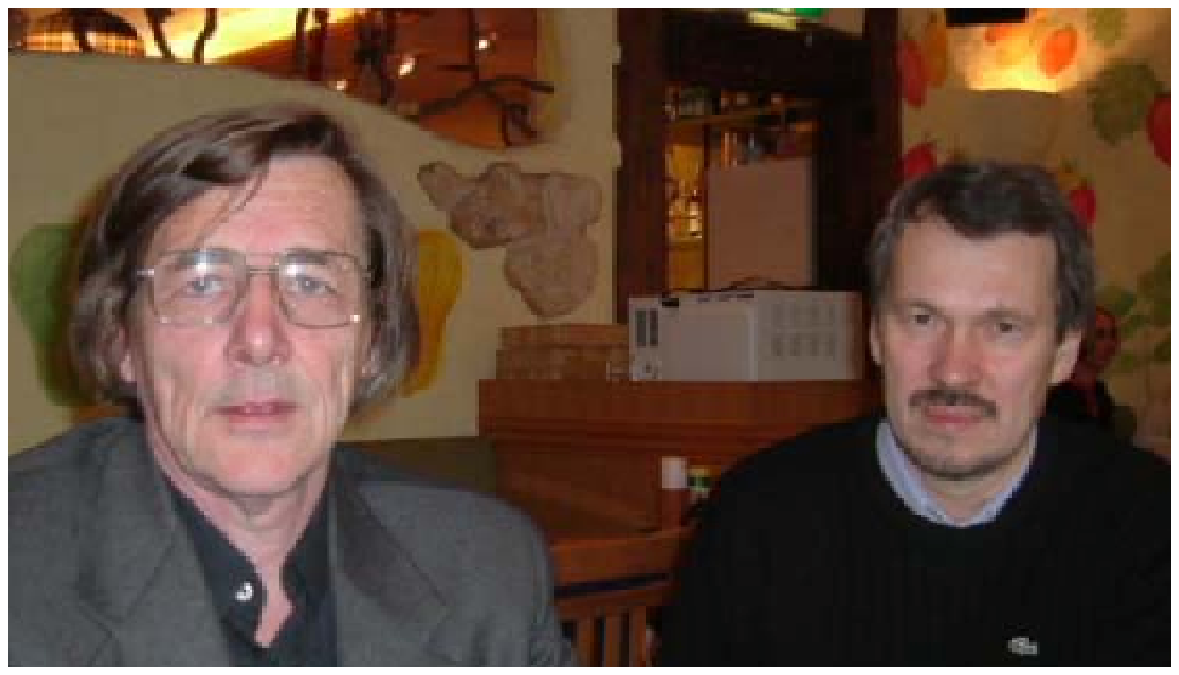}
\hfill}
\label{fig:MSW}
\caption{Lincoln Wolfenstein, Stanislav Mikheyev and Alexei Smirnov}
\end{figure}

\begin{figure}[h]   
\hspace*{1.0cm}
\includegraphics[width=6.0cm]{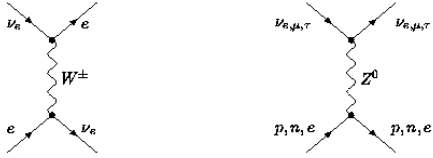}
\hspace*{2pc}
\begin{minipage}[c]{10pc}
\vspace*{-2.0cm}
\caption{\label{fig:nuint}{Neutrino interactions in matter}}
\end{minipage}   
\label{nuint}
\end{figure}

Matter effect on neutrino oscillations is due to the coherent forward 
scattering of neutrinos on the constituents of matter (fig. \ref{fig:nuint}). 
The neutral current interactions, mediated by the $Z^0$ boson exchange, are 
the same for active neutrinos of all three flavours (modulo tiny radiative  
corrections) and therefore do not affect neutrino oscillations.
In contrast to this, charged current interactions, mediated by the $W^\pm$ 
exchanges, are only possible for electron neutrinos because there 
are no muons or tauons in normal matter. This yields an effective potential 
of the electron neutrinos    
\[
~~~~~~V_{e}^{\rm CC} \equiv V = \sqrt{2}\,G_F\,N_e\,,
\]
which leads to a modification of the nature of neutrino oscillations in 
matter. 
The 2f neutrino evolution equation in matter is 
\be
\hspace*{-1.5cm}
i\frac{d}{dt}\left(\begin{array}{c}
\nu_e \\  
\nu_\mu  
\end{array} \right)=  
\left(\begin{array}{cc}
-\frac{\Delta m^2}{4E}\cos 2\theta +  V &
\frac{\Delta m^2}{4E}\sin 2\theta \\
\frac{\Delta m^2}{4E}\sin 2\theta &
\frac{\Delta m^2}{4E}\cos 2\theta
\end{array} 
\right)
\left(\begin{array}{c}
\nu_{e} \\
\nu_{\mu}
\end{array} \right)\,. 
\label{evol}
\ee
The mixing angle in matter $\theta_m$, which diagonalizes the Hamiltonian on 
the r.h.s. of eq.~(\ref{evol}), is different from the vacuum mixing 
angle $\theta$: 
\be
\sin^2 2\theta_m = \frac{\sin^2 2\theta\cdot
(\frac{\Delta m^2}{2E})^2}
{[\frac{\Delta m^2}{2E}\cos 2\theta - \sqrt{2} G_F N_e]^2+ 
(\frac{\Delta m^2}{2E})^2 \sin^2 2\theta}\,.
\ee
The flavour eigenstates can now be written as 
\be
|\nu_e\rangle ~=~ \cos\theta_m\, |\nu_{1m}\rangle +\sin\theta_m\, 
|\nu_{2m}\rangle\,,~~
\ee
\vspace*{-6mm}
\be
|\nu_\mu\rangle \,=\, -\sin\theta_m\,|\nu_{1m}\rangle +\cos\theta_m\, 
|\nu_{2m}\rangle\,,\,
\ee
where $|\nu_{1m}\rangle$ and $|\nu_{2m}\rangle$ are the eigenstates of 
the neutrino Hamiltonian in matter (matter eigenstates). The Mikheyev - 
Smirnov - Wolfenstein (MSW) resonance condition is 
\be
\sqrt{2} G_F N_e = \frac{\Delta m^2}{2E}\cos 2\theta\,.
\ee
At the resonance $\theta_m = 45^\circ$ ($\sin^2 2\theta_m=1$), i.e.  
the mixing in matter becomes maximal.  

\begin{figure}[h]   
\includegraphics[width=6.0cm,height=4.5cm]{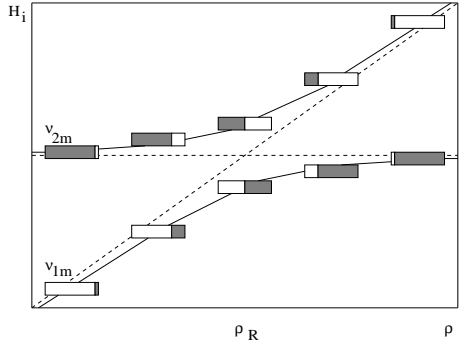}
\hspace*{6.0mm}
\begin{minipage}[c]{16pc}
\vspace*{-4.5cm}
\caption{\label{convers}
{Adiabatic neutrino flavour conversion. 
Solid curves show the energy levels of neutrino matter eigenstates,
dashed curves illustrate level crossing in the absence of mixing. 
Black and white filling corresponds to the weights of neutrino 
flavour eigenstates in given matter eigenstates.} }
\end{minipage}   
\vspace*{-2.0mm}
\end{figure}

If the matter density changes slowly enough (adiabatically) along the 
neutrino trajectory, neutrinos can undergo a flavour conversion (see fig. 
\ref{convers}). In the adiabatic regime the transitions between the matter 
eigenstates $|\nu_{1m}\rangle$ and $|\nu_{2m}\rangle$ are strongly suppressed, 
i.e. these states evolve independently. However, their flavour composition, 
which is determined by the mixing angle $\theta_m$, varies with density:
$\theta_m(N_e\gg (N_e)_{\rm res})\approx 90^\circ$, $\theta_m(N_e = 
(N_e)_{\rm res})= 45^\circ$, $\theta_m(N_e\ll (N_e)_{\rm res})\approx 
\theta$. Therefore the state produced at high densities as, e.g.,  
$\nu_e\approx \nu_{2m}$ will end up at low densities as a superposition of 
$\nu_e$ and $\nu_\mu$ with the weights $\sin^2 \theta$ and $\cos^2 \theta$, 
respectively. 
%
%
The adiabaticity (slow density change) condition can be written as 
\be
\frac{\sin^2 2\theta}{\cos 2\theta}\frac{\Delta m^2}{2E} L_\rho\gg 1\,,
\ee
where $L_\rho$ -- electron density scale height: $L_\rho = |(1/N_e)d 
N_e/d x|^{-1}$. 

A simple and useful formula for 2f conversion probability, averaged over 
production/detection positions (or small energy intervals), was derived 
in \cite{Parke}: 
\be
\overline{P}_{\rm tr}~=~\frac{1}{2}~-
~\frac{1}{2}\cos 2\theta_i \cos 2\theta_f\,(1~-~2P')\,.
\ee
Here $\theta_i$ and $\theta_f$ are the mixing angles in matter in the 
initial and final points of the neutrino path, and $P'$ is the hopping 
probability, which takes into account possible deviations from adiabaticity: 
In the adiabatic regime $P'\ll 1$, whereas in the extreme non-adiabatic 
regime $P'=\sin^2(\theta_i-\theta_f)$. 

\begin{figure}[h]
\hbox{\hfil
\hspace*{0.05cm}
\includegraphics[width=4.3cm,height=4.5cm]{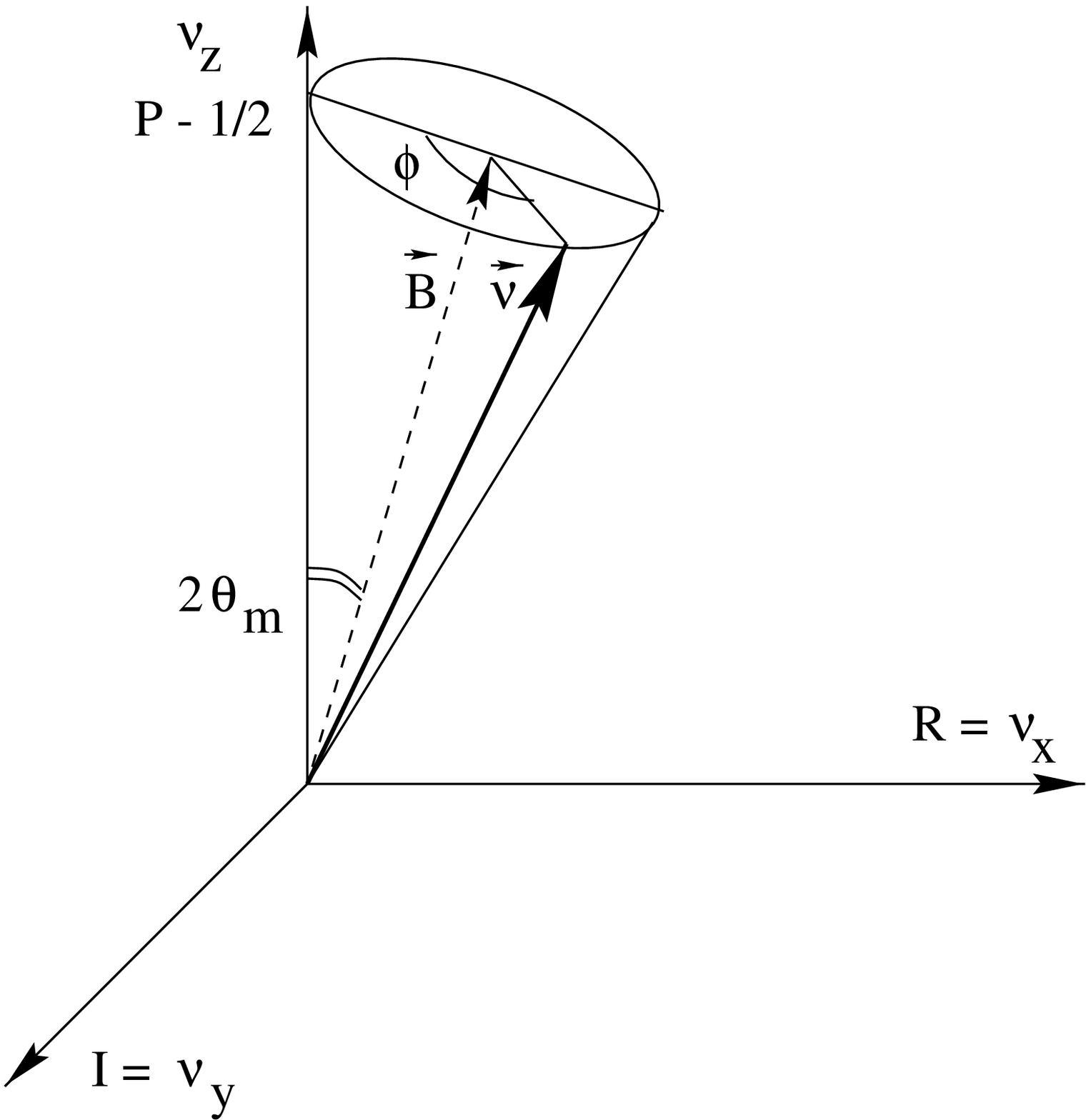}
\hspace*{0.01cm}
\includegraphics[height=4.5cm,width=3.7cm]{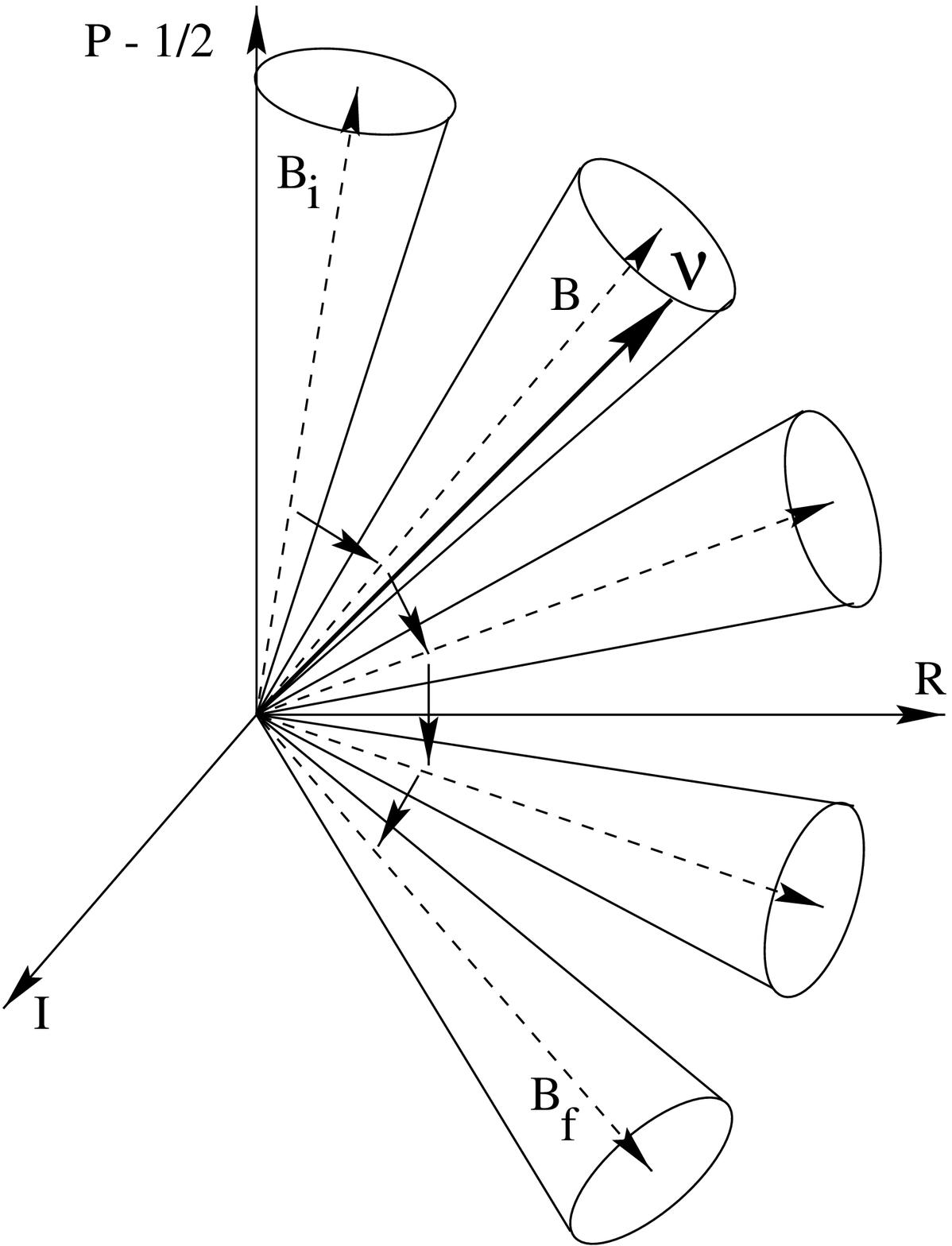}
\hspace*{0.01cm}
\includegraphics[height=4.5cm,width=3.7cm]{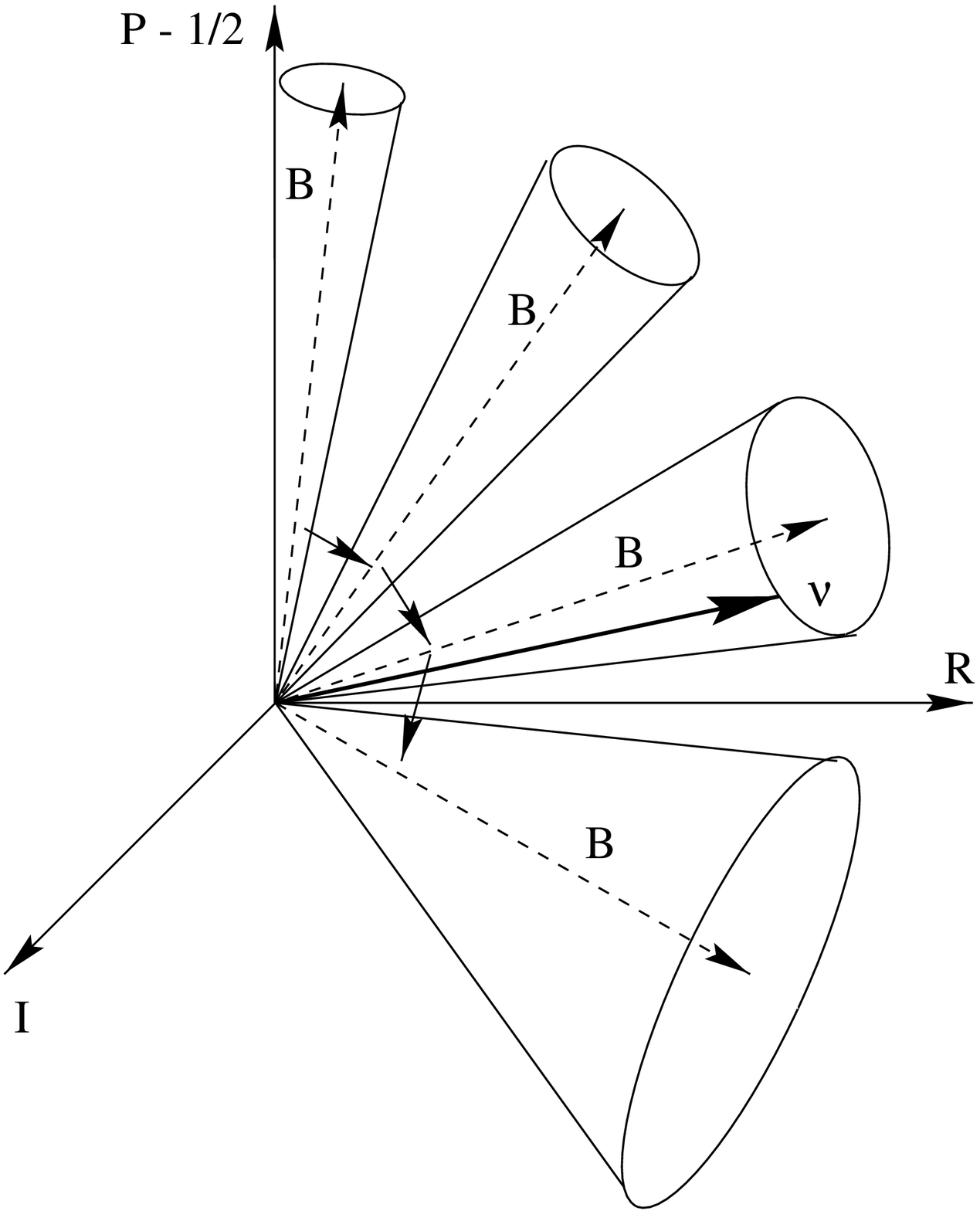}
\hfill}
\caption{\label{geom}
Analogy between neutrino oscillations in matter and spin 
precession in a magnetic field. Left panel: oscillations in constant-density 
matter, middle and right panels -- adiabatic and non-adiabatic conversions 
in matter of varying density (adopted from \cite{Sm1}). }
\end{figure}

The evolution equation for the neutrino system can be also written as  
\be
\frac{d\vec{S}}{dt} = 2 (\vec{B}\times\vec{S})\,,\qquad 
\mbox{where}\qquad
\vec{S} = \{{\rm Re}(\nu_e^*\nu_\mu)\,,~{\rm Im}(\nu_e^*\nu_\mu)\,,
~\nu_e^* \nu_e-1/2\}\,,
\ee
\be
\vec{B} = \{(\Delta m^2/4E)\sin 2\theta_m\,,~~0\,, ~~V/2-(\Delta m^2/4E)
\cos 2\theta_m\}\,.
\ee
The first equation here coincides with the equation for spin precession in 
a magnetic field. This analogy can be used for a graphical illustration of 
neutrino oscillations in matter (see fig. \ref{geom}).

\begin{figure}[h]   
\includegraphics[width=3.5cm]{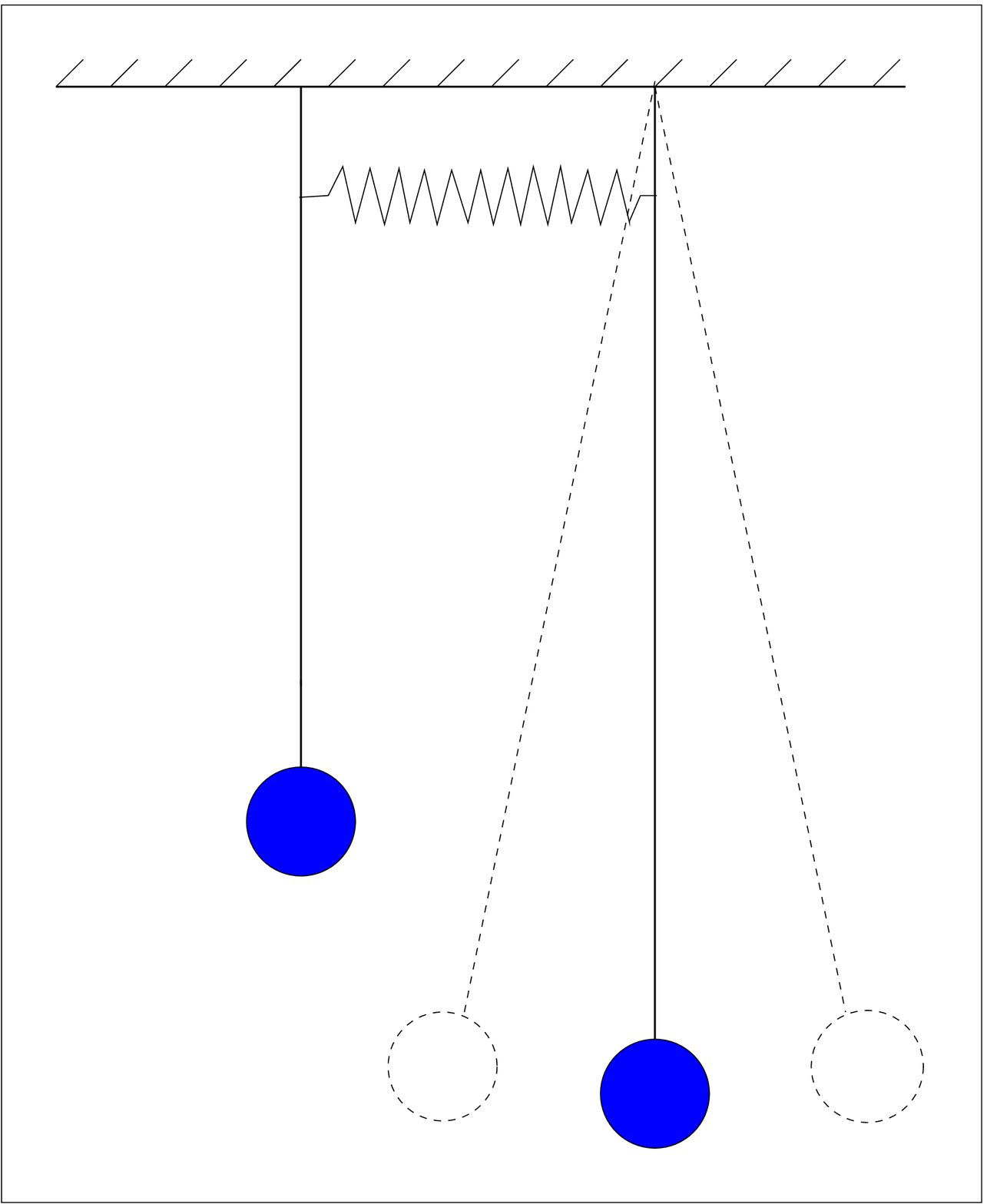}
\hspace*{0.4cm}
\includegraphics[width=3.5cm]{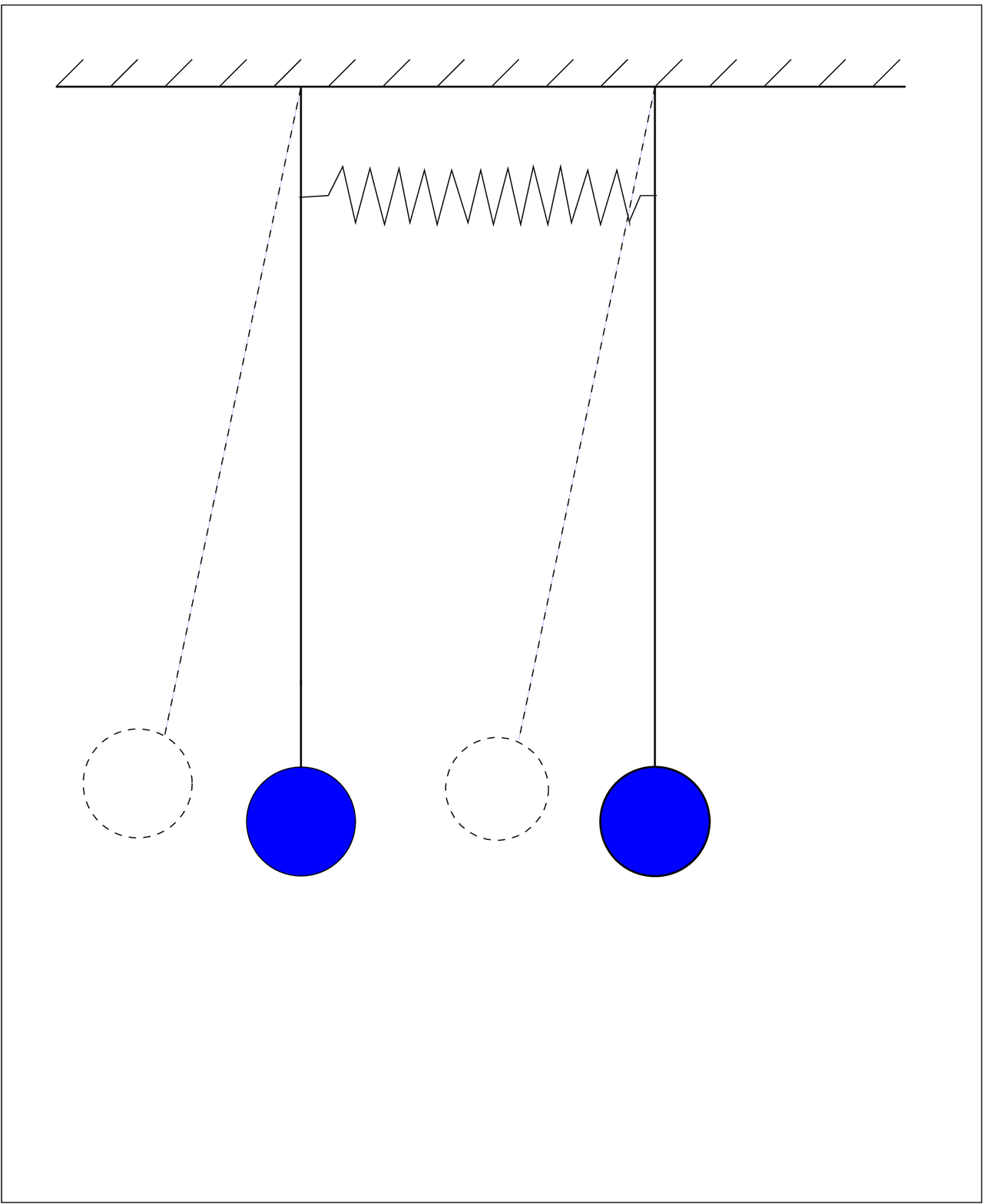}
\hspace*{0.4cm}
\includegraphics[width=3.5cm]{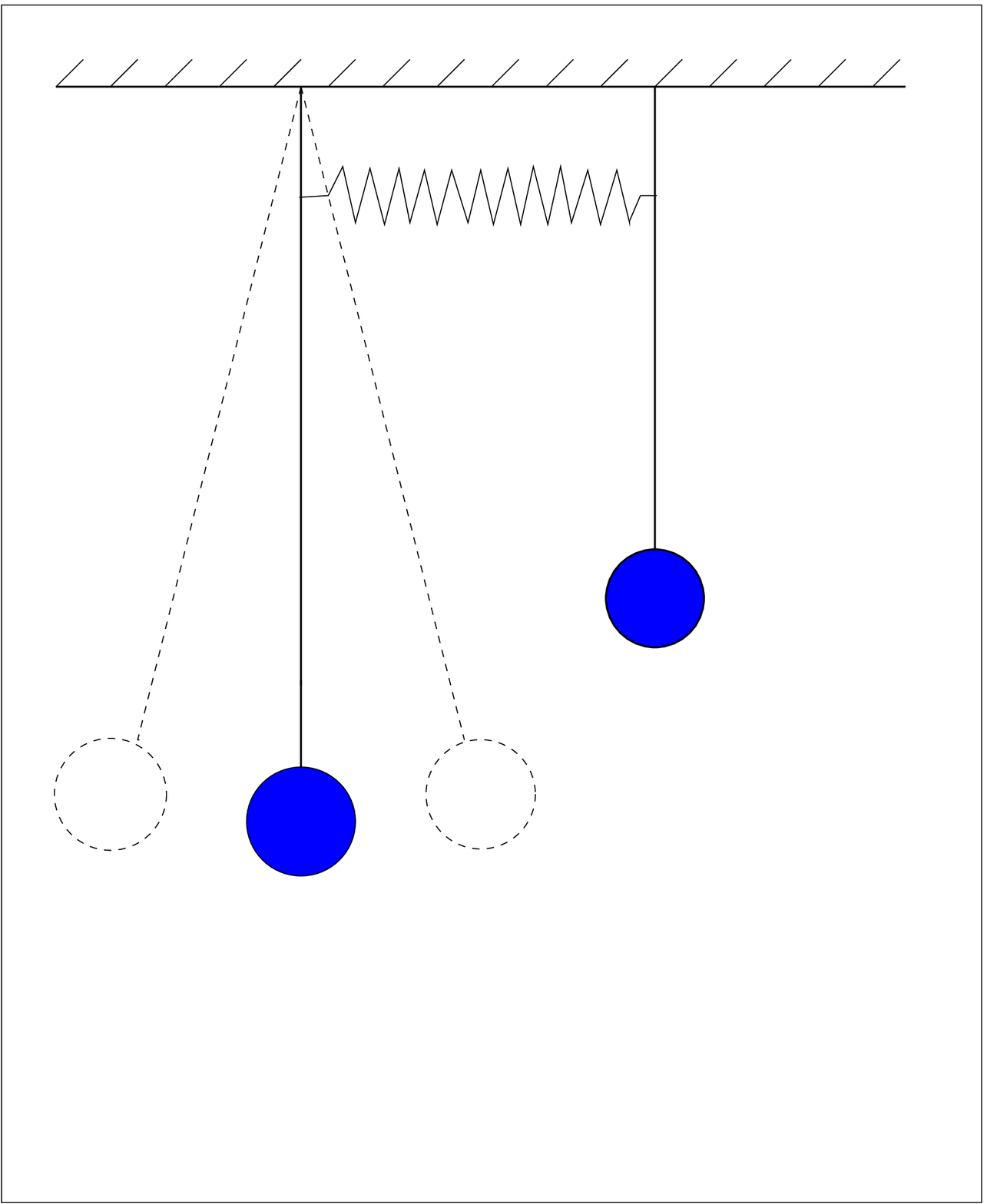}
\hspace*{0.8pc}
\begin{minipage}[c]{8pc}
\vspace*{-3.5cm}
\caption{\label{mechan}
Mechanical analogue of neutrino flavour conversion  
in matter -- two coupled pendula of variable lengths. }
\end{minipage}   
\end{figure}

Another analogy of neutrino flavour conversion in matter is provided by a 
system of two coupled pendula \cite{pend} (see fig. \ref{mechan}). When the 
right pendulum gets a kick, it starts oscillating, but the left pendulum is 
almost at rest because the eigenfrequencies of the two pendula are very 
different. With the length $l_2$ of the right pendulum slowly decreasing, 
its eigenfrequency approaches that of the left one, and when $l_2=l_1$ 
the two frequencies coincide (the resonance occurs): both pendula oscillate 
with the same amplitude. When the length of the right pendulum decreases 
further, the amplitude of its oscillations decreases too, while the left 
pendulum starts oscillating with a large amplitude. This adiabatic transfer of 
the oscillation energy from one pendulum to another is analogous to the 
adiabatic neutrino flavour conversion.

\begin{figure}[h]
\includegraphics[width=4.8cm,angle=270]{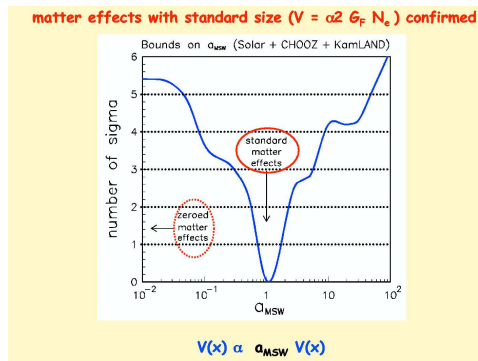}
\hspace*{2pc}
\begin{minipage}[t]{16pc}
\vspace*{1.0cm}
\caption{\label{MSWresults}
Results of the analysis of the solar, CHOOZ and KamLAND data 
with the standard matter-induced potential rescaled by a factor $a_{MSW}$, 
treated as a free parameter.  The value $a_{\rm MSW} \approx 1$ is strongly 
favoured \cite{Bari}. }
\end{minipage}   
\end{figure}

Analysis of the solar neutrino data and the results of the KamLAND and CHOOZ 
reactor neutrino experiments has convincingly demonstrated that the (large 
mixing angle) MSW effect is responsible for the flavour conversion of solar 
neutrinos, thus resolving the long-standing problem of the deficiency 
of the observed flux of solar neutrinos. This is illustrated by the analysis 
of the Bari group, in which the strength of the matter-induced 
potential of electron neutrinos was considered a free parameter 
(fig. \ref{MSWresults}). For more on MSW effect, see the talk of 
A. Friedland \cite{Fried}.

\section{Phenomenology}

All the available neutrino data except those of the LSND experiment can 
be explained in terms of oscillations between the 3 known neutrino species 
-- $\nu_e$, $\nu_\mu$ and $\nu_\tau$. If the LSND results are correct, they 
would most likely require the existence $\ge 1$ light sterile neutrinos 
$\nu_s$ (though some exotic scenarios also exist: CPT violation, violation of 
Lorentz invariance, mass-varying neutrinos, shortcuts in extra dimensions, 
decaying neutrinos, etc.). 
The MiniBooNE experiment was designed to confirm or refute the LSND claim, 
and the results are expected very soon. One should remember, however, that 
even if the LSND result is not confirmed, this would not rule out the 
existence of light sterile neutrinos and ~$\nu_{\rm a}\leftrightarrow 
\nu_{\rm s}$ oscillations, which is an intriguing possibility with important  
implications for particle physics, astrophysics and cosmology. From now 
on I will concentrate on 3-flavour (3f) oscillations of active neutrinos. For 
more on sterile neutrinos, see the talk of A. Kusenko \cite{Kus}.

\subsection{3-flavour neutrino mixing and oscillations}

For 3 neutrino species the  mixing matrix depends in general on 3 mixing 
angles $\theta_{12}$, $\theta_{23}$ and $\theta_{13}$, one Dirac-type 
CP-violating  phase $\delta_{\rm CP}$, and two Majorana-type CP 
violating phases $\sigma_{1,2}$. The Majorana-type phases can be factored 
out in the mixing matrix according to $U=U_0 K$, 
$K = {\rm diag}(1\,, e^{i \sigma_1}\,, e^{i \sigma_2})$. The factor $K$ 
does not affect neutrino oscillations and so will be omitted hereafter.  
Renaming $U_0\to U$, the relevant part of the leptonic mixing matrix can be 
written in the standard parameterization as 
\be
U = 
\left( 
\begin{array}{ccc}
c_{12} c_{13} &
 s_{12} c_{13} & s_{13} e^{-i \delta_{\rm CP}} \\ -s_{12}
 c_{23} - c_{12} s_{13} s_{23} e^{i \delta_{\rm CP}}
& c_{12} c_{23} - s_{12} s_{13} s_{23} e^{i \delta_{\rm CP}} & c_{13}
s_{23} \\ s_{12} s_{23} - c_{12} s_{13} c_{23} e^{i \delta_{\rm CP}} &
-c_{12} s_{23} - s_{12} s_{13} c_{23} e^{i \delta_{\rm CP}} & c_{13}
c_{23} 
\nonumber
\end{array}
\right)\,,
\ee
where 
$s_{ij}=\sin\theta_{ij}$ and $c_{ij}=\cos\theta_{ij}$. 

Neutrino oscillations probe the neutrino mass squared differences, which 
satisfy $\Delta m_{\rm sol}^2\equiv \Delta m_{21}^2\ll \Delta m_{32}^2
\simeq \Delta m_{31}^2\equiv \Delta m_{\rm atm}^2$. Accordingly, 
there are two possible orderings of the neutrino masses: normal hierarchy, 
when the mass eigenstate $\nu_3$, separated from $\nu_1$ and $\nu_2$ by the 
largest mass gap, is the heaviest one, and inverted hierarchy, when 
$\nu_3$ is the lightest state (see fig. \ref{hier}). 

\vspace*{7mm}
\hspace*{12mm}  
Normal hierarchy: \hspace*{2.8cm} Inverted hierarchy:
\vspace*{3mm}

\begin{figure}[h]
\epsfxsize=5.5cm\epsfbox{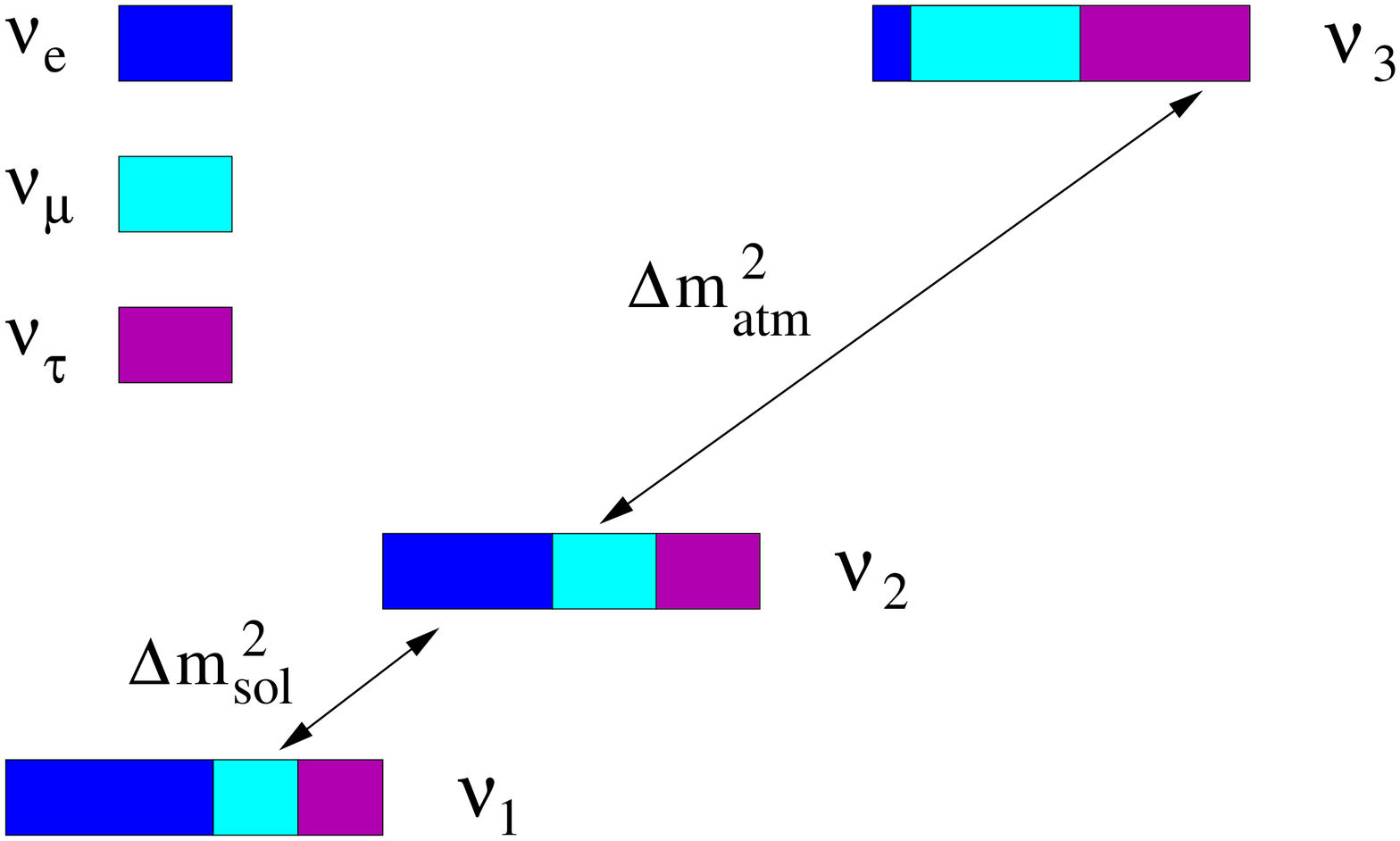}
\hspace{4mm}
\epsfxsize=5.55cm\epsfbox{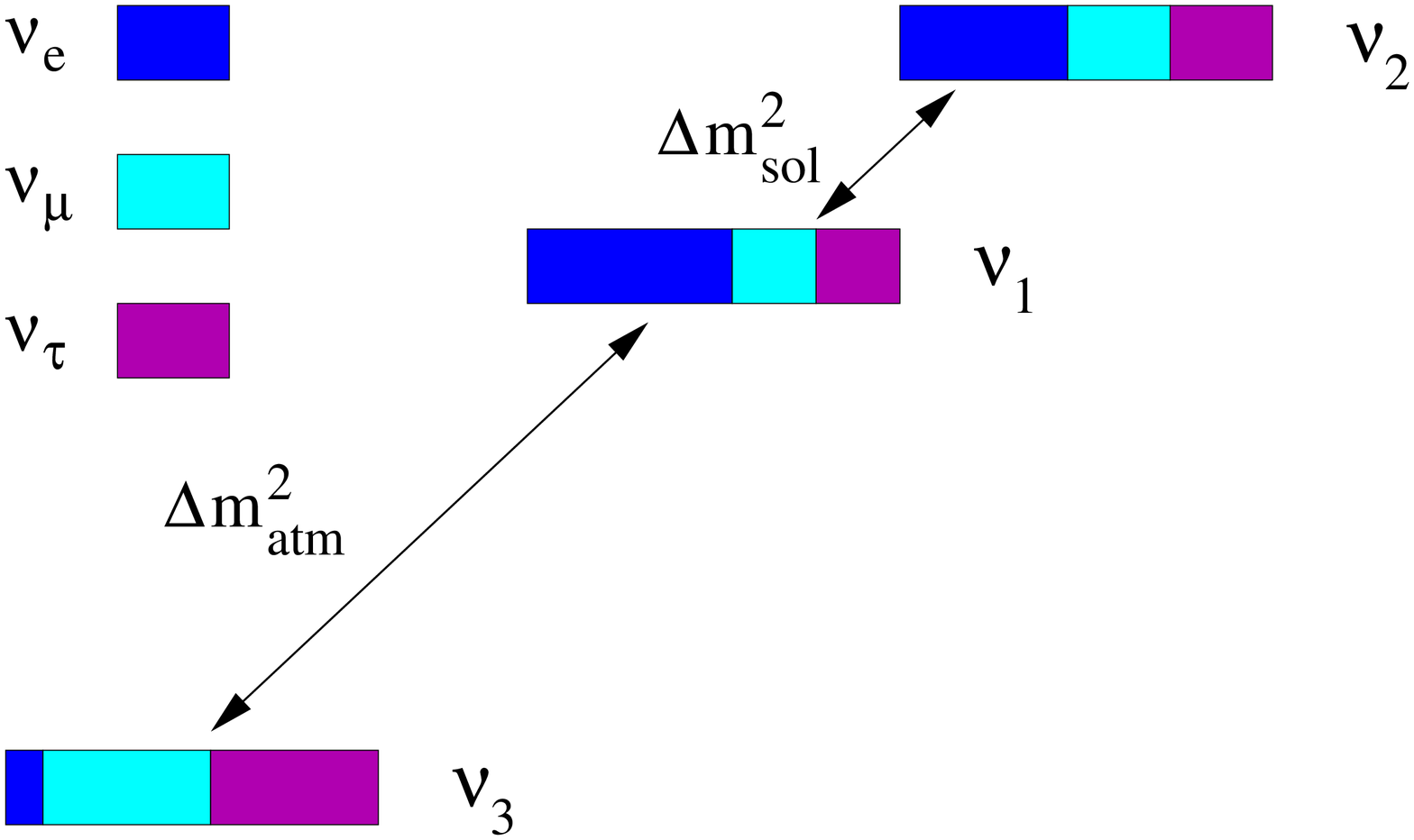}
\hspace*{1pc}
\begin{minipage}[c]{8pc}
\vspace*{-3.5cm}
\caption{\label{hier}
Normal and inverted neutrino mass orderings. The different 
fillings show the relative weights of different flavour eigenstates in 
given mass eigenstates. } 
\end{minipage}   
\end{figure}

\subsection{2f and effective 2f approximations}

In many cases 2f description of neutrino oscillations gives a good first
approximation. The reasons for this are   
(i) the hierarchy of $\Delta m^2$: $\Delta m_{\rm sol}^2 \ll \Delta m_{\rm 
atm}^2$, and (ii) the smallness of ~$|U_{e3}|$. 
There are exceptions, however: when oscillations due to the solar frequency 
($\propto \Delta m_{\rm sol}^2$) are not frozen, the probabilities $P(\nu_\mu 
\leftrightarrow \nu_\tau)$, $P(\nu_\mu \rightarrow \nu_\mu)$ and $P(\nu_\tau 
\rightarrow \nu_\tau)$ do not have a 2f form \cite{series}. However, even 
for the probabilities of oscillations involving $\nu_e$, the corrections due 
to 3-flavourness can be as large as $\sim 10\%$, i.e. are at the same level as 
the accuracy of the present-day data, and so cannot be ignored. In addition, 
there is a number of very interesting pure 3f effects in neutrino oscillations. 
Therefore, 3f analyses are now a must.

For oscillations driven by $\Delta m_{\rm sol}^2$, the third neutrino mass 
eigenstate $\nu_3$ essentially decouples. However, there is still a ``memory'' 
of this state through unitarity, which results in the emergence of 
various powers of $c_{13}$ in the expressions for transition 
probabilities. An example is the survival probability of solar $\nu_e$ 
\cite{Lim} (the same expression also applies for the survival probability 
of reactor $\bar{\nu}_e$ observed in KamLAND) : 
\be
P_D(\nu_e\to \nu_e) ~\simeq~  c_{13}^4 P_{2ee}(\Delta m_{21}^2,
~\theta_{12}, 
~c_{13}^2 V) ~+~ s_{13}^4\,.\qquad
\label{solar}
\ee
The term $s_{13}^4$ is tiny and can be safely neglected. Another example is the 
day-night effect for solar $\nu_e$:  while the day-time survival probability 
$P_D(\nu_e) \propto c_{13}^4$, the difference of the night-time and day-time 
probabilities  $P_N(\nu_e)-P_D(\nu_e)\propto c_{13}^6$ \cite{BOS,ATV}. 
Deviations from 2f expressions (which correspond to the limit $\theta_{13}\to 
0$) may be substantial: for the maximal currently experimentally allowed 
values of $\sin^2 2\theta_{13}$ one has $(1-c_{13}^4)\simeq 0.1$, 
$(1-c_{13}^6)\simeq 0.13$.

For oscillations of reactor $\bar{\nu}_e$, the survival probability can to 
a very good accuracy be written as 
\be P_{\bar{e}\bar{e}} \simeq 1-\sin^2 2\theta_{13} \cdot \sin^2
\left(\frac{\Delta m_{31}^2}{4E} L\right) -c_{13}^4 \sin^2 2\theta_{12} 
\cdot \sin^2\left(\frac{\Delta m_{21}^2}{4E}L\right) \,.
\label{reac}
\ee 
Since the average energies of reactor antineutrinos $\bar{E}\sim 4$ MeV, 
for experiments with relatively short baseline ($L\lesssim 1$ km), such as 
CHOOZ, Palo Verde and Double CHOOZ, one has 
$(\Delta m_{21}^2/4E)\,L \ll 1$.  Eq. (\ref{reac}) then reduces to 
\be
P(\bar{\nu}_e\to\bar{\nu}_e; L) = 1- \sin^2 2\theta_{13} \cdot
\sin^2\left(\frac{\Delta m_{31}^2}{4E} \;L\right)\,,
\ee
i.e. takes the 2f form. Note that the ``solar'' term $\sim \sin^2 
2\theta_{12}$ in (\ref{reac}) cannot be neglected if $\theta_{13} \lesssim 
0.03$, which is about the reach of currently discussed future reactor 
experiments. 

For the unique long-baseline reactor neutrino experiment KamLAND ($\bar{L}
\simeq 170$ km) one has $(\Delta m_{21}^2/4E)\,L \gtrsim 1$, $(\Delta 
m_{31}^2/4E)\,L\gg 1$, and the $\bar{\nu}_e$ survival probability takes 
the effective 2f form as in eq.~(\ref{solar}). 
Note that matter effects in KamLAND should be of order a few per cent,  
i.e. can be comparable with 3f corrections due to $\theta_{13}\ne 0$. 

\subsection{Genuine 3f effects in neutrino oscillations}
 
These are, first of all, CP and T violation. CP violation results 
in $P(\nu_a\to \nu_b)\ne P(\bar{\nu}_a\to \bar{\nu}_b)$, whereas T violation 
leads to $P(\nu_a\to \nu_b)\ne P(\nu_b\to \nu_a)$. 
Under the standard assumptions of locality and normal relation between 
spin and statistics, quantum field theory conserves CPT.  
CPT invariance of neutrino oscillations in vacuum gives $P(\nu_a\to\nu_b) 
= P(\bar{\nu}_b \to\bar{\nu}_a)$. Therefore CP violation implies T 
violation 
and vice versa.

One can consider the following probability differences as measures of CP 
and T violation:  
\be
\Delta P_{ab}^{\rm CP} \equiv P(\nu_a\to \nu_b)-P(\bar{\nu}_a\to 
\bar{\nu}_b)\,,\qquad 
\Delta P_{ab}^{\rm T}\, \equiv\, P(\nu_a\to \nu_b)-P(\nu_b\to\nu_a)\,.
\ee
{}From CPT invariance, for oscillations in vacuum one has
\be
\Delta P_{ab}^{\rm CP} = \Delta P_{ab}^{\rm T}\,, 
\quad\quad\quad\quad
\Delta P_{aa}^{\rm CP} = 0\,.
\ee
In the 3f case there is only one Dirac-type CP-violating phase 
$\delta_{\rm CP}$ and therefore only one CP and T violating probability 
difference:
\be
\Delta P_{e\mu}^{\rm CP} ~=~ \Delta P_{\mu\tau}^{\rm CP} ~=~
\Delta P_{\tau e}^{\rm CP} ~\equiv~ \Delta P\,, \quad \mbox{where}
\ee
\[
\Delta P ~=~ {} -\, 4 s_{12}\,c_{12}\,s_{13}\,c_{13}^2\,s_{23}\,c_{23}\,
\sin\delta_{\rm CP}
\quad \quad \quad \quad \quad \quad \quad \quad \quad \quad \quad
\]
\be
\quad\quad\quad \times \left[\sin\left(\frac{\Delta m_{12}^2}{2E}\,L
\right)+ \sin\left(\frac{\Delta m_{23}^2}{2E}\,L\right)+ \sin\left(
\frac{\Delta m_{31}^2}{2E}\,L\right)\right]\,.
\ee
This probability difference vanishes when one or more of the following 
conditions are satisfied:  
at least one $\Delta m_{ij}^2 = 0$; 
at least one $\theta_{ij} = 0$ or $90^\circ$; 
$\delta_{\rm CP}=0$ or $180^\circ$; 
in the regime of complete averaging; 
in the limit $L\to 0$ ($\Delta P\to 0$ as $L^3$).
Obviously, the effects of CP and T violating are very difficult to observe!
For more on that, see the talk of O. Mena \cite{Mena}.

{\em CP violation and T violation in $\nu$ oscillations in matter}.   
Normal matter (with number of particles $\ne$ number of antiparticles)  
violates C, CP and CPT, which leads to a fake (extrinsic) CP violation in 
neutrino oscillations. It exists even in the 2f limit and may complicate 
the study of the fundamental (intrinsic) CP violation. 

The situation with T-violation in matter is different: matter with density 
profile symmetric w.r.t. the midpoint of neutrino trajectory does not 
induce any fake T violation. Asymmetric profiles do, but only for $N>2$ 
flavors \cite{deGouv,Tviol}. Matter-induced T violation is an interesting pure 
3f effect; it may fake fundamental T violation and complicate its study 
(extraction of $\delta_{\rm CP}$ from the experiment). However, it is absent 
when either $U_{e3}=0$ or $\Delta m_{\rm sol}^2=0$ (2f limits) and thus is 
doubly suppressed by both these small parameters. Therefore its effects in 
terrestrial experiments are expected to be very small \cite{Tviol}.

\begin{figure}[h]
\includegraphics[width=4.8cm]{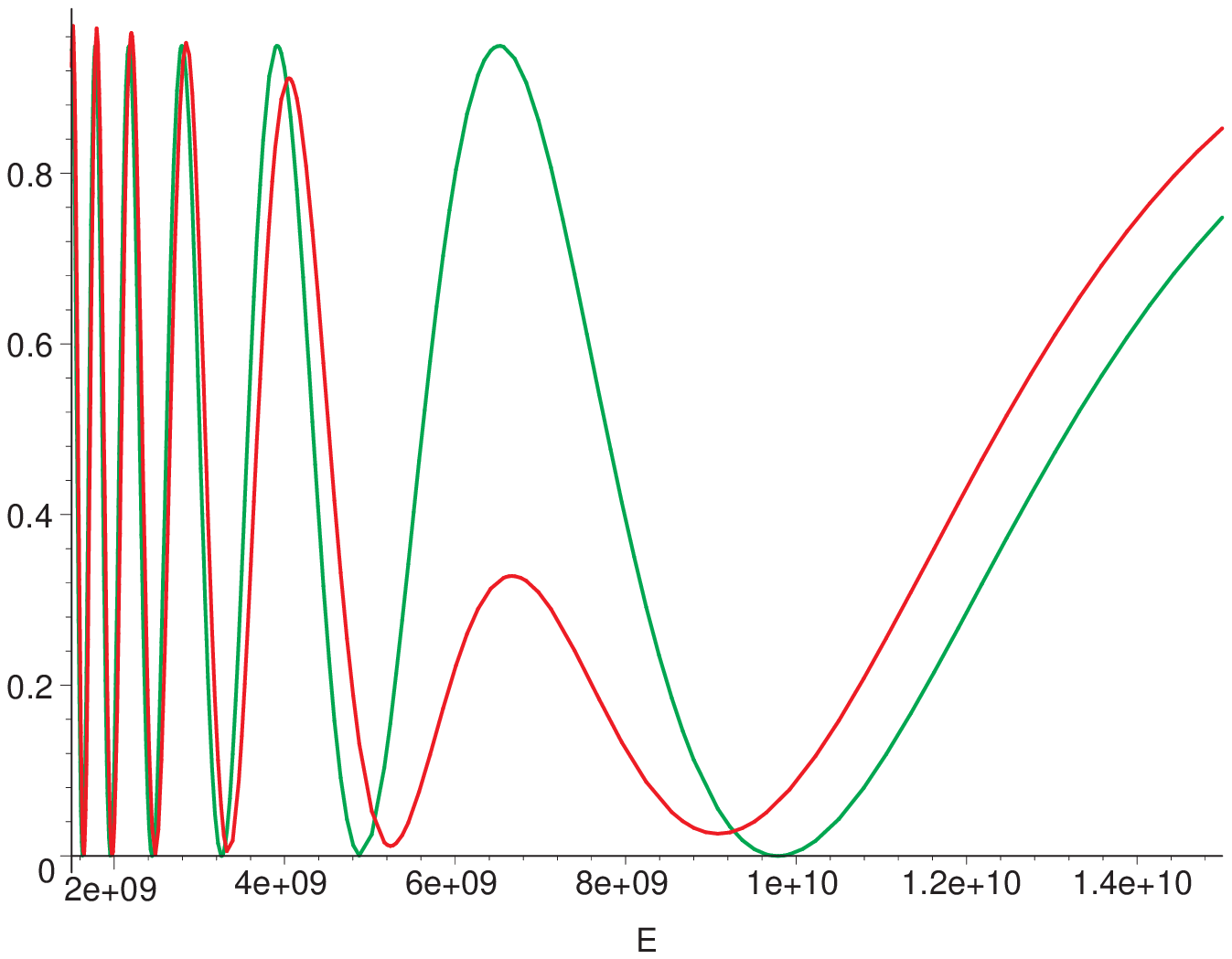}
\hspace*{0.2cm}
\includegraphics[width=4.8cm]{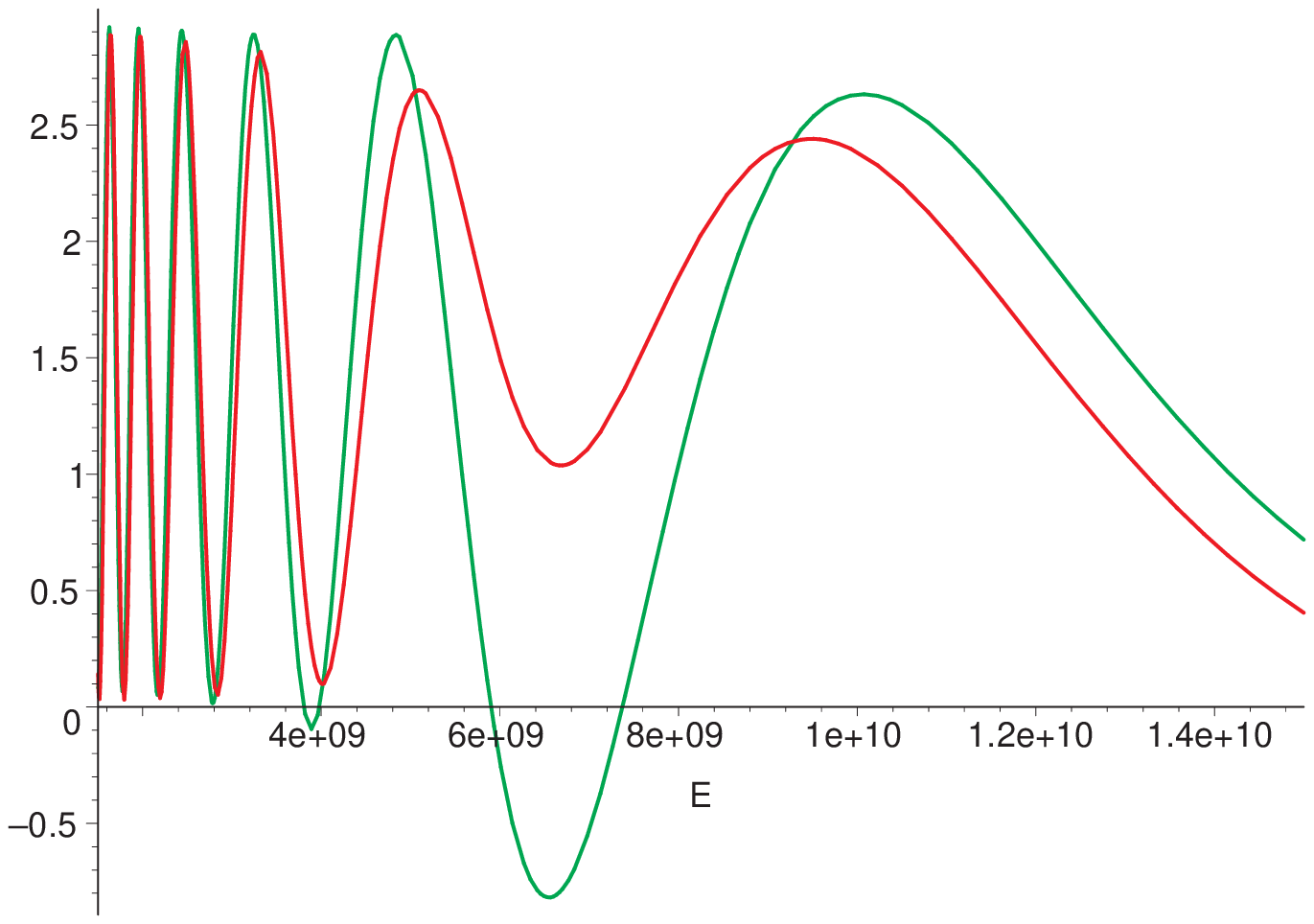}
\hspace*{0.8pc}
\begin{minipage}[c]{12pc}
\vspace*{-3.7cm}
\caption{\label{matteff} Left panel: $P_{\mu\tau}$. Right panel: oscillated 
flux of atmospheric $\nu_\mu$. $\Delta m_{31}^2=2.5\times 10^{-3}$ {\rm 
eV}$^2$, $\sin^2\theta_{13}=0.026$, $\theta_{23}=\pi/4$, $\Delta m_{21}^2=0$, 
$L=9400$ {\rm km}. Red (dark) curves -- with matter effects, green (light) 
curves -- without matter effects on $P_{\mu\tau}$. }
\end{minipage}   
\end{figure}

{\em Matter effects on $\nu_\mu \leftrightarrow \nu_\tau$ oscillations}. 
In the 2f limit, matter does not affect $\nu_\mu \leftrightarrow \nu_\tau$ 
oscillations (because the matter-induced potentials $V(\nu_\mu)$ and 
$V(\nu_\tau)$ coincide up to tiny radiative corrections). However, this 
is not true in the full 3f framework \cite{3fnumunutau}. In particular, for 
oscillations inside the earth there are ranges of baselines and neutrino 
energies for which the matter effect can be very large (fig. \ref{matteff}, 
left panel, $E\sim 5$ -- 10 GeV). If one ignores them, one may end up 
with a negative expected flux of oscillated $\nu_\mu$ in atmospheric neutrino 
experiments (fig. \ref{matteff}, right panel). 

\subsection{Parametric resonance in neutrino oscillations in matter}

\begin{figure}[h]
\hspace*{0.3cm}
\includegraphics[width=3.0cm]{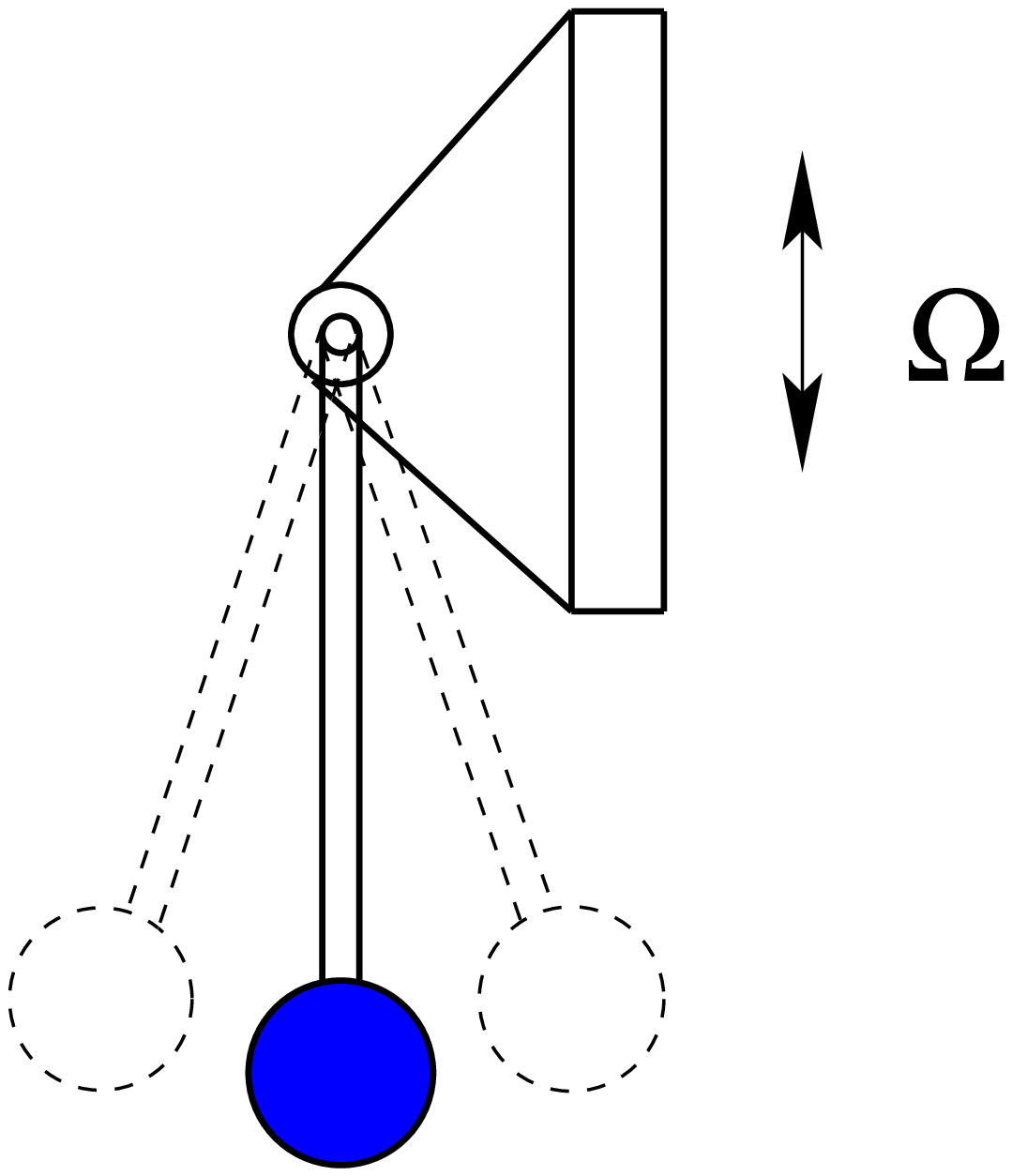}
\hspace*{0.3cm}
\includegraphics[width=3.0cm]{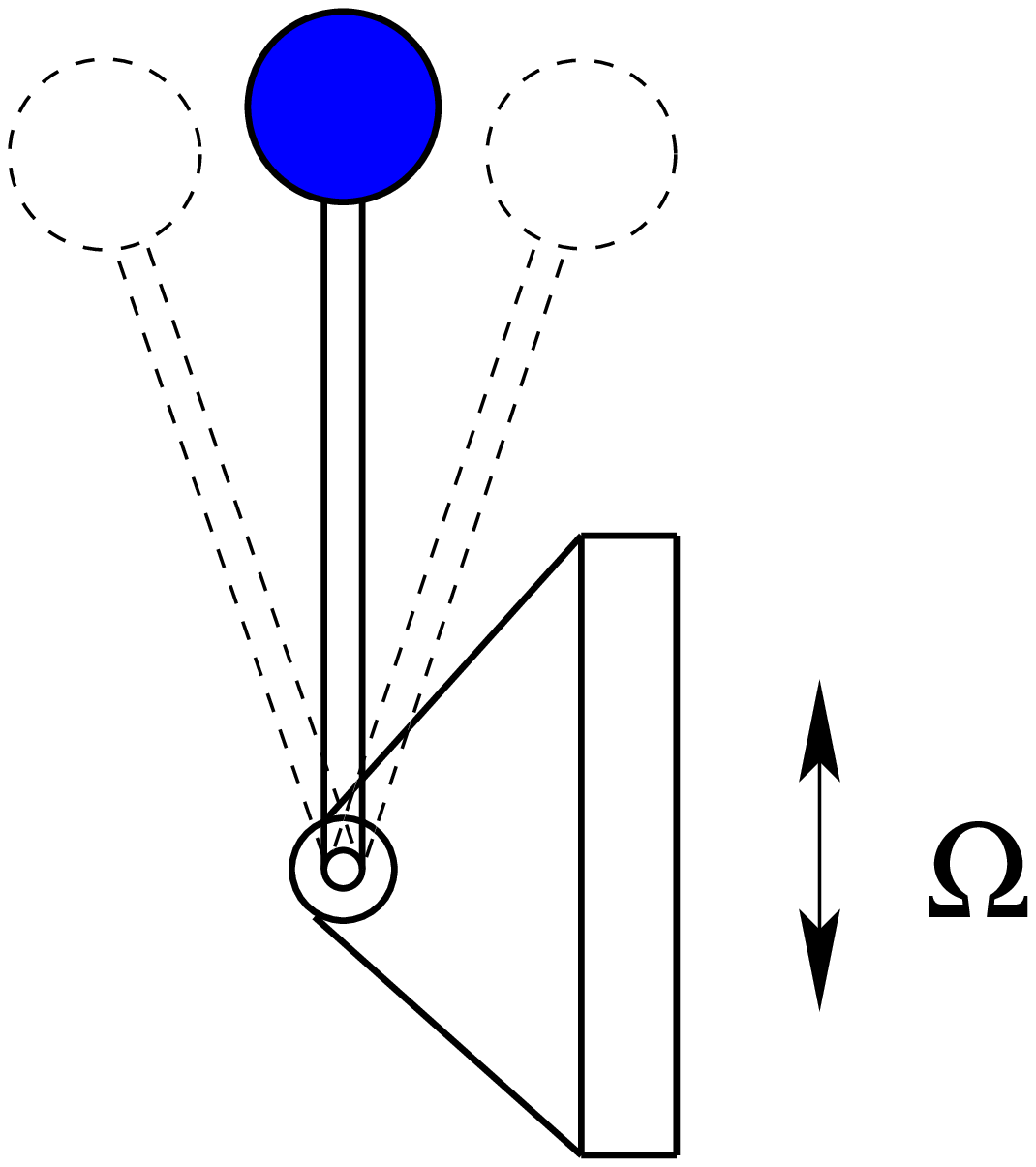}
\hspace*{1.0cm}
\begin{minipage}[c]{18pc}
\vspace*{-3.5cm}
\caption{\label{parammech} Parametric resonance in oscillations of a pendulum 
with vertically oscillating point of support. For small-amplitude oscillations 
the resonance condition is $\Omega_{\rm res} = 2\omega/n$ $(n=1, 2, 3...)$.
}
\end{minipage}   
\end{figure}

\begin{figure}[h]
\hbox{\hfil
\includegraphics[width=7.8cm,height=4.5cm]{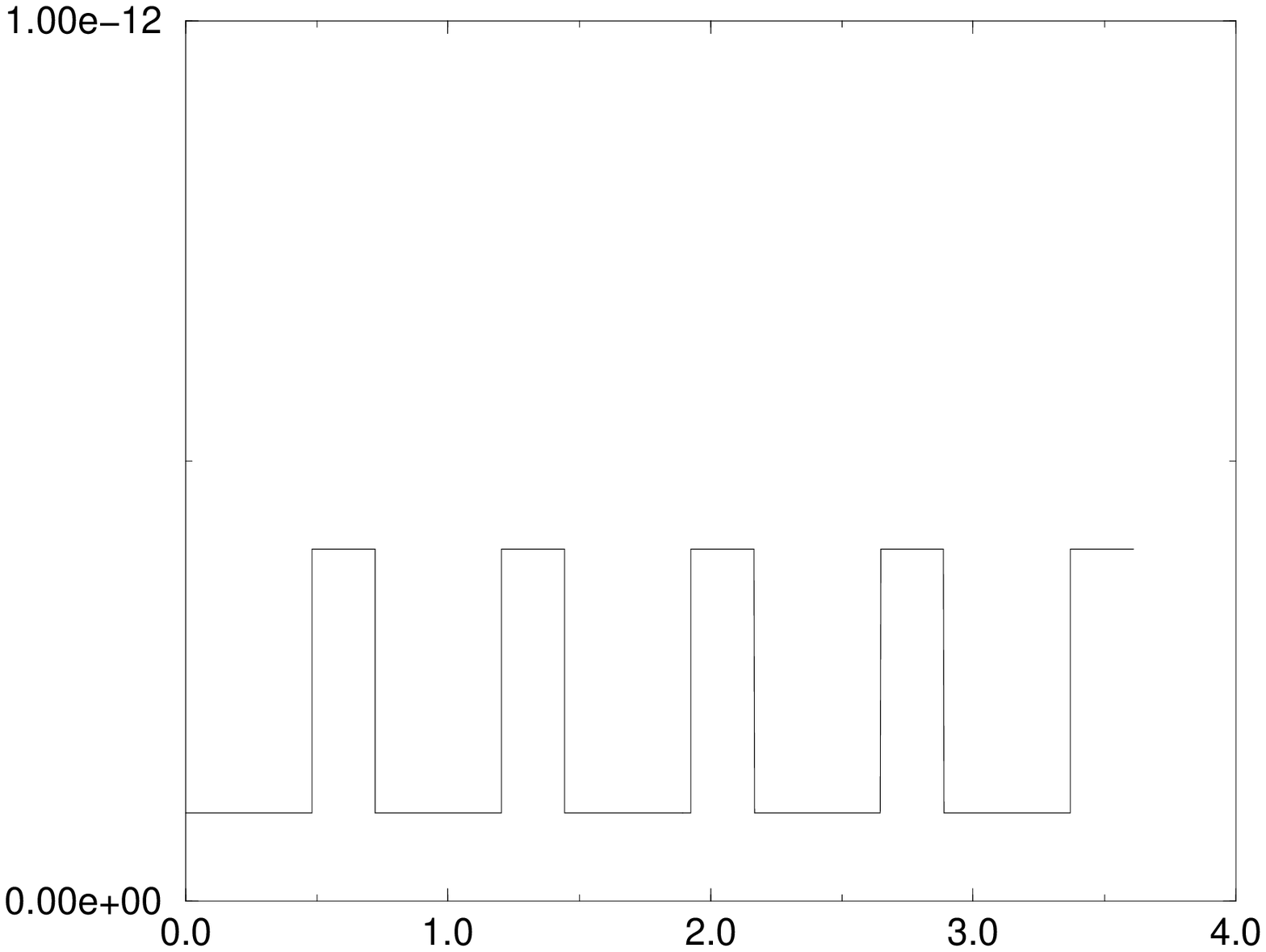}
\includegraphics[width=7.8cm,height=4.5cm]{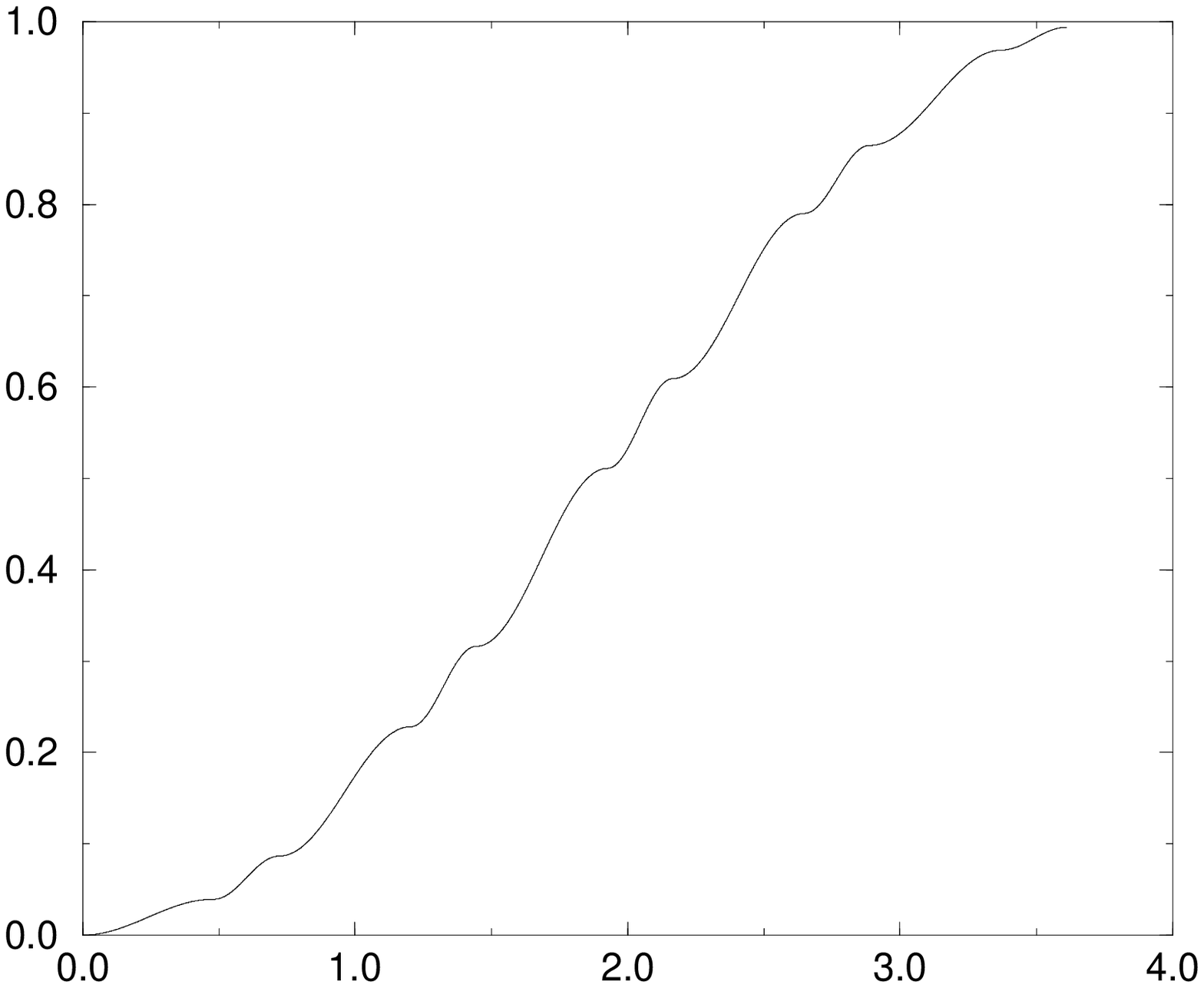}
\hfill}
\caption{\label{fig:paramres} Parametric resonance in the case of a ``castle 
wall'' density profile. 
Coordinate dependence of the potential $V$ (left panel) 
and of the transition probability $P$ (right panel). }
\end{figure}

The MSW effect is not the only possible way matter can influence neutrino 
oscillations. Another interesting possibility is a parametric enhancement of 
neutrino oscillations in matter \cite{ETC,Akhm1}. Parametric resonance in 
oscillating systems with varying parameters occurs when the rate of the 
parameter change is correlated in a certain way with the values of the 
parameters themselves. A well-known mechanical example is a pendulum with 
vertically oscillating suspension point (fig. \ref{parammech}): when the 
frequency $\Omega$ and amplitude $A$ of these oscillations are in a special 
correlation with the eigenfrequency $\omega$ and amplitude $a$ of the pendulum, 
the pendulum can turn upside down and start oscillating around the vertical, 
normally unstable, equilibrium point. 
Neutrino oscillations in matter can undergo parametric enhancement if the 
length and size of the density modulation is correlated in a certain way with 
neutrino parameters. This enhancement is completely different from the MSW 
effect; in particular no level crossing is required. An example admitting an 
exact analytic solution is the ``castle wall'' density profile 
\cite{Akhm1,Akhm2} (see fig. \ref{fig:paramres}). 
The resonance condition in this case can be written as \cite{Akhm2} 
\be
X_3\equiv -(\sin\phi_1 \cos\phi_2\cos 2\theta_{1m}+\cos\phi_1
\sin\phi_2\cos 2\theta_{2m}) = 0\,,
\label{X3}
\ee
where $\phi_{1,2}$ are the oscillation phases acquired in layers 1 and 2 
and $\theta_{m 1,2}$ are the corresponding mixing angles in matter. 

\begin{figure}[h]
\hspace*{0.005cm}
{\includegraphics[width=6.2cm]{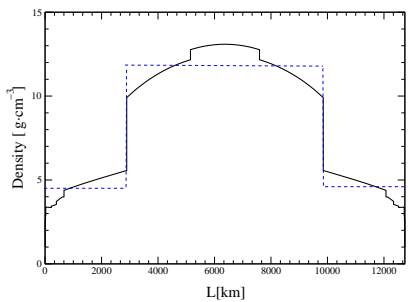}}
\hspace*{0.4cm}
\begin{minipage}[c]{20pc}
\vspace*{-3.5cm}
\caption{\label{prem} The earth density profile according to the PREM model 
\cite{PREM} and its approximation by a piece of the ``castle wall'' profile. }
\end{minipage}   
\end{figure}

\begin{figure}[h]
\hspace*{-0.4cm}
\includegraphics[width=6.4cm,height=4.8cm]{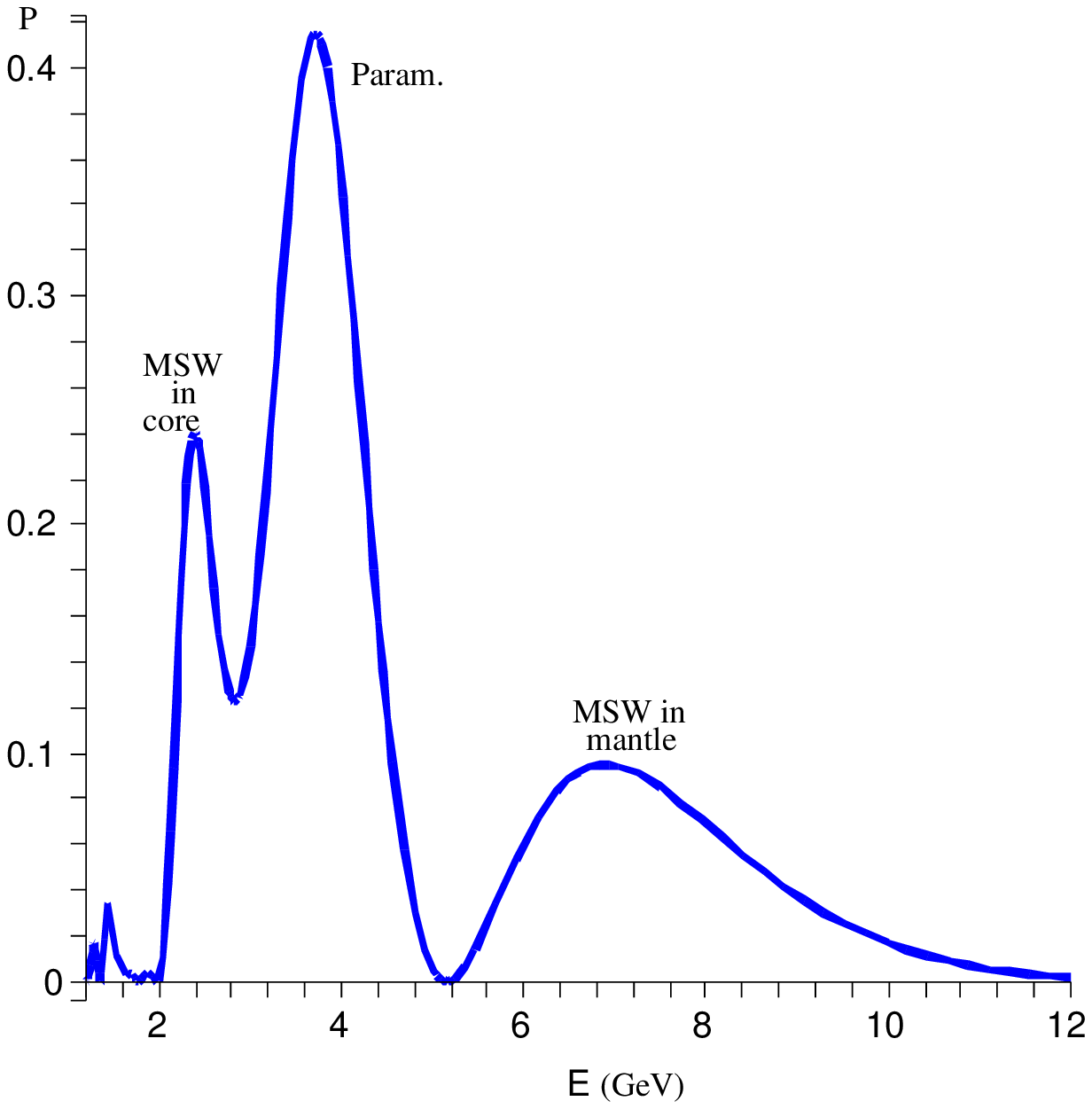}
\hspace*{-0.95cm}
\includegraphics[width=5.9cm,height=4.8cm]{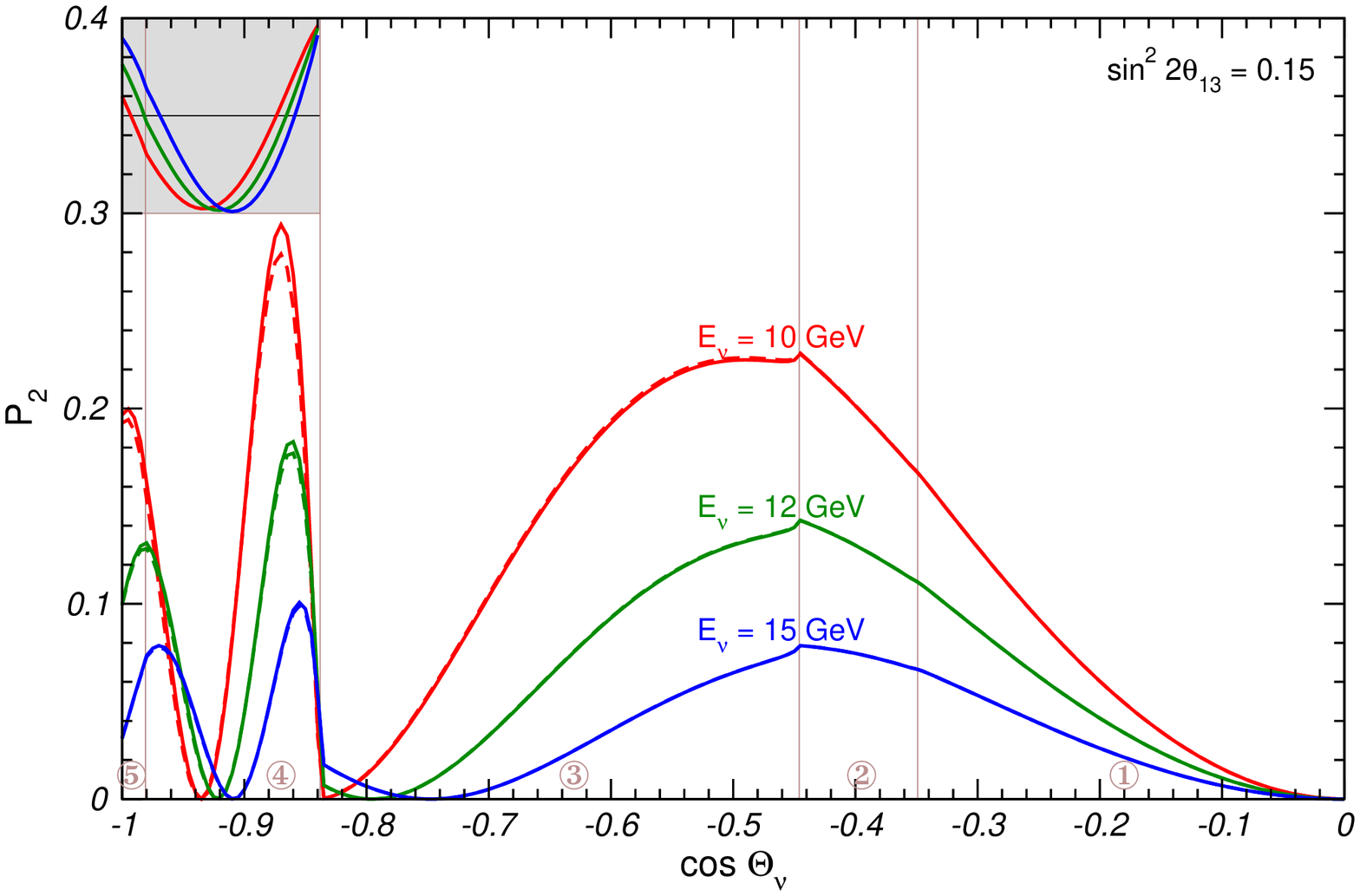}
\hspace*{0.2cm}
\begin{minipage}[c]{10pc}
\vspace*{-4.4cm}
\caption{\label{illustr} Parametric resonance in oscillations of core-crossing 
neutrinos. Left panel: $\cos\Theta=-0.89$, $\sin^2 2\theta_{13}=0.01$; right 
panel: $\sin^2 2\theta_{13}=0.15$, small window shows the values of $X_3$.
}
\end{minipage}   
\end{figure}

\begin{figure}[h]
\includegraphics[width=9.8cm]{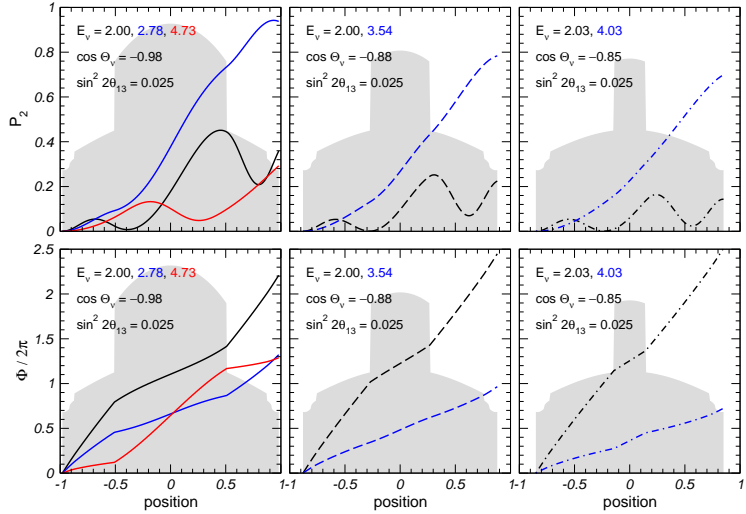}
\hspace*{0.3cm}
\begin{minipage}[c]{14pc}
\vspace*{-4.6cm}
\caption{\label{coord} Coordinate dependence of the transition 
probability $P_2=1-P_{ee}$ (upper panels) and of the total adiabatic 
oscillation phase $\Phi$ (lower panels) for a number of values of $E$ and 
$\cos\Theta_\nu$. The shaded areas show the earth's density profiles seen 
by the neutrinos. Figure courtesy of M. Maltoni. 
}
\end{minipage}   
\end{figure}

The earth's density profile seen by neutrinos with core-crossing 
trajectories can be well approximated by a piece of this castle wall 
profile (fig.~\ref{prem}). 
Interestingly, the parametric resonance condition (\ref{X3}) can be satisfied 
for oscillations of core-crossing neutrinos in the earth for a rather wide 
range of zenith angles both at intermediate energies \cite{LS,P,Akhm2} and 
high energies \cite{AMS1} (see figs. \ref{illustr}, \ref{coord}). 
%
The parametric resonance of neutrino oscillations in the earth can be 
observed in future atmospheric or accelerator experiments if $\theta_{13}$ is 
not too much below its current upper limit.

\subsection{Some recent developments}
When $V \ll \Delta m^2/2E$ (oscillations of low-$E$ neutrinos in matter or, 
equivalently, oscillations in low-density matter), matter effects on neutrino 
oscillations are small and can be considered in perturbation theory. This 
gives simple and transparent formulas describing, in particular, oscillations 
of solar and supernova neutrinos in the earth. The earth matter effects can be 
expressed through the regeneration factor $f_{reg}=P_{2e}^{\oplus}-
P_{2e}^{vac}$, where $P_{2e}$ is the probability for $\nu_2$ to become $\nu_e$ 
upon traversing the earth. In the 3f framework one has \cite{ATV} 
\be
P_{2e}^\oplus-P_{2e}^{vac}=\frac{1}{2}\,c_{13}^4\,\sin^2
2\theta_{12}\,\int_0^L\! dx \,V(x) \sin\left[2\int\limits_x^{L}\!\omega(x')
\,dx'\right]\,,
\label{P2e}
\ee
where 
\be
\omega(x)=\sqrt{[\cos 2\theta_{12}\,\delta-c_{13}^2 V(x)/2]^2+\delta^2
\sin^2 2\theta_{12}}\,,\qquad \delta=\frac{\Delta m_{21}^2}{4E}\,.
\label{omega}
\ee
The 2f ($\theta_{13}=0$) version of these equations was derived in \cite{IS} 
(see also \cite{dHLS}).  

\begin{figure}[h]
\hbox{\hfill
\includegraphics[width=8cm,height=4.5cm]{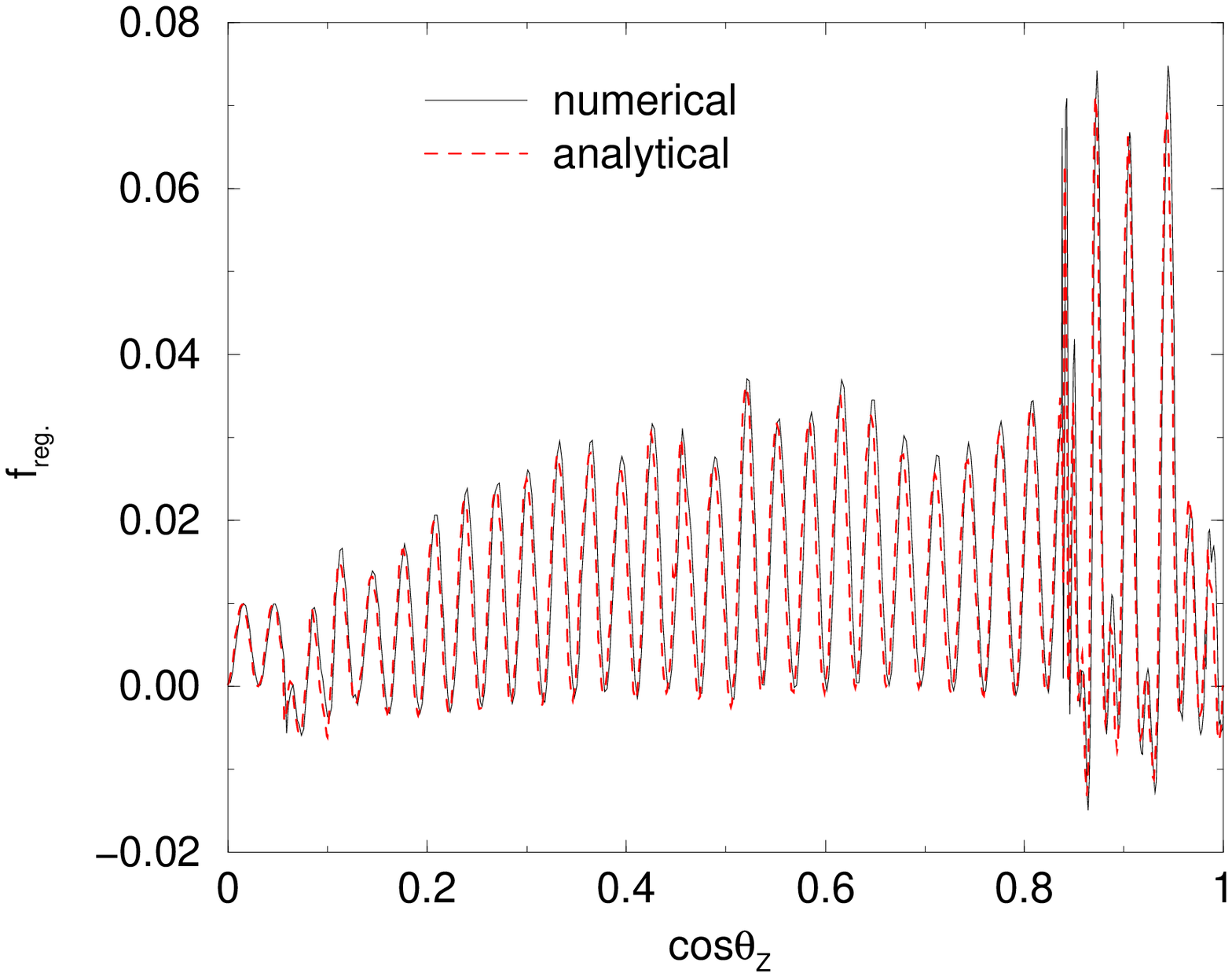}
\includegraphics[width=8cm,height=4.5cm]{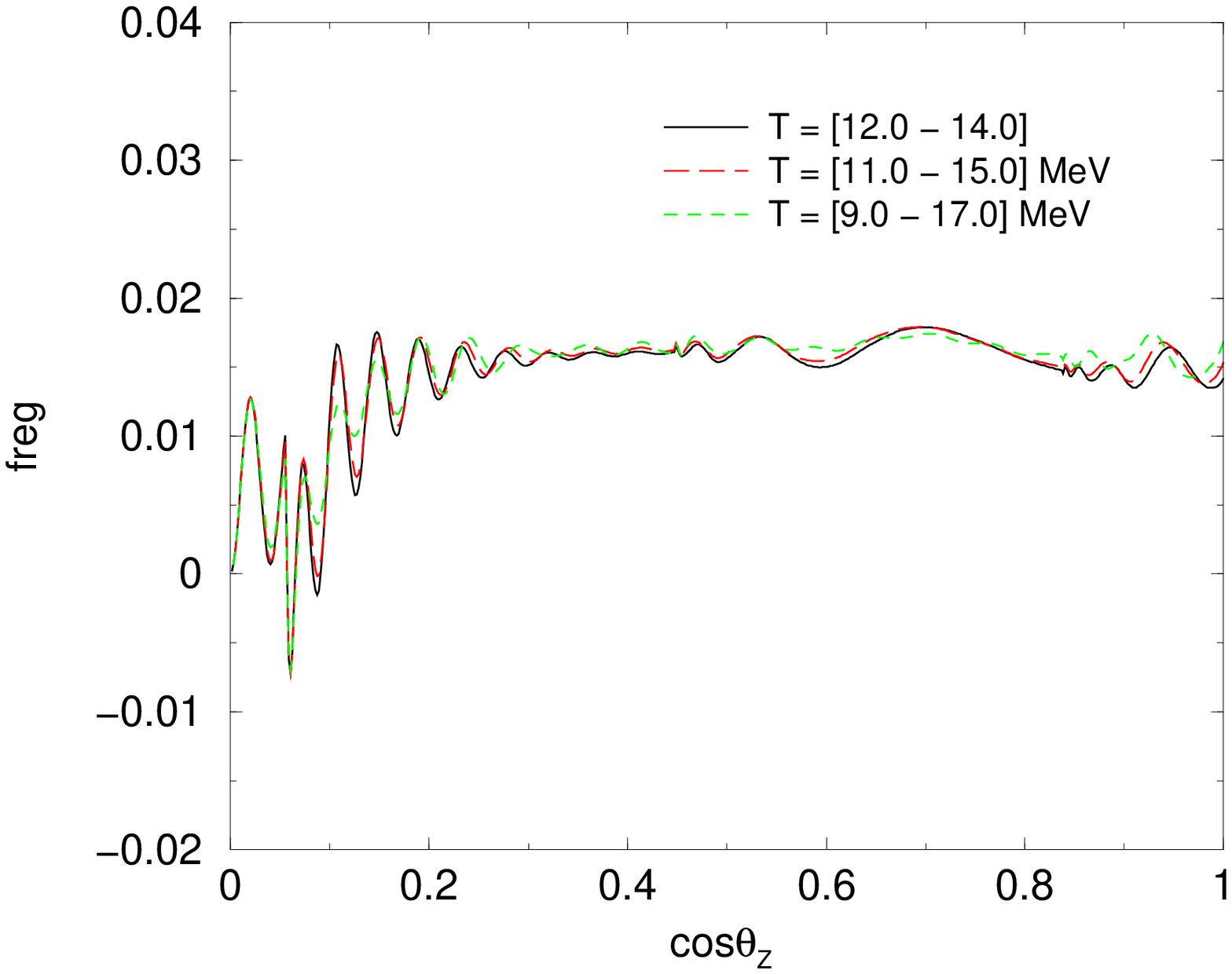}
\hfill}
\caption{\label{reg} Left panel: earth regeneration factor for $E=10$ MeV 
neutrinos, perfect energy resolution. Black (solid) curve -- numerical 
calculation, red (dashed) curve -- analytic result. Right panel: regeneration 
factors averaged over three different intervals of energy \cite{dHLS}. }
\end{figure}

The regeneration factor as the function of the cosine of the nadir angle 
$\Theta_z$ is shown in fig. \ref{reg}. As can be seen in the left panel, in 
the case of perfect energy resolution one could expect a significant increase 
of the regeneration factor for core-crossing trajectories. However, 
experimentally no such an increase was observed in the $\cos\Theta_z$ 
dependence of the day-night signal difference for solar neutrinos, which is 
rather flat. As was shown in \cite{IS}, this comes about because of the 
finite energy resolution $\Delta E$ of the detectors, which leads to a 
suppression of the effects of the earth density variations that are far from 
the detector (see the right panel of fig. \ref{reg}), the attenuation length 
being $d\simeq l_{\rm osc}(E/\Delta E)$.

Oscillations of high energy neutrinos in matter or, equivalently, oscillations 
in dense matter ($V > \delta \equiv\Delta m^2/4E$), can also be very 
accurately described analytically. The transition probability for oscillations 
in a matter of an arbitrary density profile is given by \cite{AMS1} 
\be
P~=~\delta^2 \sin^2 2\theta\left|\int_0^L dx e^{-2 i \phi(x)}\right|^2,
\qquad \quad
\phi(x)=\int_0^x dx' \omega(x')\,.
\label{highE}
\ee
The accuracy of this approximation is quickly increasing with neutrino 
energy (see the right panel of fig. \ref{illustr}, where the exact results 
are shown by solid curves and the analytic results, by dashed curves). 
Eq.~(\ref{highE}) also allows a simple analytic interpretation of the two 
prominent parametric peaks in the core region, seen in this figure \cite{AMS1}.

\section{Unsettled issues?}

The theory of neutrino oscillations is quite mature and well developed now. 
However, it is far from being complete or finished, and a number of basic 
questions are still being debated. 
Below I list some of these  
questions (given in italics), along with my short answers to them: 
\footnote{
Detailed discussion could not be given for the lack of time.} 
\begin{itemize} 

\item 
{\em Equal energies or equal momenta? }

-- Neither equal $E$ nor equal $p$ assumptions normally used in the 
derivations of the oscillation \hspace*{2.5mm}probability are exact. But for 
relativistic neutrinos, both give the correct answer.

\item 
{\it Evolution in space or in time? }

-- This is related to the previous question. For relativistic neutrinos both 
are correct and \hspace*{2.5mm} equivalent. Fortunately, in practice we 
only deal with relativistic neutrinos. In the non-\hspace*{2.5mm} relativistic 
case the very notion of the oscillation probability is ill-defined (the 
probability \hspace*{2.5mm} depends on both the   
production and detection processes).  

\item
{\em Claim: evolution in time is never observed}

-- Incorrect. ~Examples: ~K2K, MINOS (and now also CNGS) experiments, which 
use the \hspace*{2.5mm} neutrino time of flight in order to suppress the 
background. 

\item 
{\em Is wave packet description necessary? }

-- Yes, if one wants to rigorously justify the standard oscillation 
probability formula. Once \hspace*{2.5mm} this is done, the wave 
packets can be forgotten unless the issues of coherence become \hspace*{2.5mm} 
important.

\item 
{\em Do charged leptons oscillate? }

-- No, they don't. 

\item 
{\em Is the standard oscillation formula correct? }

-- Yes, it is. In particular, there is no extra factor of two in the 
oscillation phase, which \hspace*{2.5mm} is sometimes claimed to be there. 
However, it would be theoretically interesting and \hspace*{2.5mm} important 
to study the limits of applicability of the standard approach.  

\end{itemize} 
\vspace*{2mm}

A number of subtle issues of the neutrino oscillation theory still remain 
unsettled (e.g., rigorous wave packet treatment, oscillations of 
non-relativistic neutrinos, etc). At present, this is (rightfully) of little 
concern for practitioners.

What are interesting future tasks for the theory and phenomenology of neutrino 
oscillations?  These include the search for the best strategies for measuring 
neutrino parameters, study of subleading effects and effects of non-standard 
neutrino interactions and of the domains of applicability and limitations of 
the current theoretical framework. Future experimental results may also bring 
some new surprises and pose more challenging problems for the theory!

\ack 
The author was supported by the Wenner-Gren 
Foundation as the Axel Wenner-Gren visiting professor at the Royal 
Institute of Technology.

\end{document}